\documentclass[aps,prd,notitlepage,groupedaddress,nofootinbib,superscriptaddress]{revtex4-1}
\newcommand{\be}{\begin{eqnarray}}
\newcommand{\ee}{\end{eqnarray}}
\usepackage{fancyhdr}
\usepackage{amsmath}
\usepackage{mathrsfs}
\usepackage{float}
\usepackage{xcolor}
\usepackage{graphicx, epsfig, amssymb} 
\usepackage{amsmath, amsfonts}
\usepackage{bm} 
\usepackage{tensor}
\usepackage{soul}
\usepackage{hyperref}
\usepackage{physics}

\newcommand{\com}{{\cal C}}

\begin{document}

\title{Analytical thresholds for black hole formation in general cosmological backgrounds}
\author{Albert Escriv\`a}
\email{albert.escriva@fqa.ub.edu}
\affiliation{Institut de Ci\`encies del Cosmos, Universitat de Barcelona, Mart\'i i Franqu\`es 1, 08028 Barcelona, Spain}
\affiliation{Departament de F\'isica Qu\`antica i Astrof\'isica, Facultat de F\'isica, Universitat de Barcelona, Mart\'i i Franqu\`es 1, 08028 Barcelona, Spain}
\email{albert.escriva@fqa.ub.edu}
\author{Cristiano Germani}
\email{germani@icc.ub.edu}
\affiliation{Institut de Ci\`encies del Cosmos, Universitat de Barcelona, Mart\'i i Franqu\`es 1, 08028 Barcelona, Spain}
\affiliation{Departament de F\'isica Qu\`antica i Astrof\'isica, Facultat de F\'isica, Universitat de Barcelona, Mart\'i i Franqu\`es 1, 08028 Barcelona, Spain}
\author{Ravi K. Sheth}
\email{shethrk@upenn.edu}
\affiliation{Center for Particle Cosmology, University of Pennsylvania, Philadelphia, PA 19104, USA}

\begin{abstract}
  We consider black holes which form from an initially spherically symmetric super-Hubble perturbation of a cosmological background filled by a perfect fluid $p=w \rho$ with $w\in (0,1]$.  
  Previous work has shown that when $w = 1/3$ (radiation), there is a critical threshold for black hole formation ($\delta_c$), which, to a very good approximation, only depends upon the curvature of the compaction function around its peak value.  We find that this generalizes to all $w\gtrsim 1/3$; for smaller $w$s the knowledge of the full shape of the compaction function is necessary. We provide analytic approximations for $\delta_c$ which are accurate for $w\in [1/3,1]$.  
\end{abstract}

\maketitle

\section{Introduction}\label{sec:intro}

Primordial Black Holes (PBHs), first theorized in \cite{Zeldovich,hawking1,hawking2,Carr:1975qj,Novikov}, could have formed in the very early Universe from the gravitational collapse of cosmological perturbations. Several estimates of the PBHs abundance suggest that they may make up a significant fraction, if not all, of the Dark Matter (DM) today \cite{last}.

PBH formation is studied by considering the evolution of initially super-Hubble perturbations.  The simplest ones are spherically symmetric, and are characterized by the way in which their ``compaction function'' (roughly the``gravitational potential'') varies with scale $r$.  The compaction function generically has a maximum on some scale $r=r_m$.  A PBH forms if, on this scale, the compaction function exceeds a certain critical threshold $\delta_c$.  The predicted statistical abundances of PBHs typically depend strongly on the value of this threshold.  For example, in the case of PBH formation during a radiation-dominated epoch, the abundance is exponentially sensitive to $\delta_c$ (e.g. \cite{meilia, jaume-yoo, ravi-cri}).  

Early estimates  of $\delta_c$ (e.g. \cite{carr75} and \cite{harada}) were based on simplified analytically solvable models under certain rather restrictive assumptions. These were used to motivate the existence of a ``universal'' threshold that was supposed to apply for any equation of state.  However, numerical studies have shown that, even for a fixed equation of state, $\delta_c$ is not universal \cite{musco2005,hawke2002,refrencia-extra-jaume,Niemeyer2,nakama2,sasaki,EAG, ilia, albert_paper}.  The main reason is that $\delta_c$ depends on the details of the initial perturbation \cite{ilia}, i.e., on the scale dependence or ``shape'' of the compaction function. Nevertheless, it was shown in \cite{RGE} that during a radiation-dominated epoch (equation of state $p=w\rho$ with $w=1/3$), to a very good approximation, there exists a universal (shape independent) threshold value for the volume-averaged compaction function.  Since the volume average is dominated by scales near the maximum of the compaction function, in \cite{RGE} we showed that it is sufficient to parameterize the profile dependence of $\delta_c$ by the curvature of the compaction function at its maximum. Using this insight, we found an analytic approximation to the shape dependence of $\delta_c$ which matches that found in simulations to within a few percent.

This raises the question of whether or not this universality is generic.  There are at least two directions to explore: non-spherical perturbations, and equations of state for the background that differ from radiation.

The critical threshold required to form a black hole from an a-spherical configuration is generally larger than for the spherical case \cite{florian}. The reason is very simple: in a-spherical configurations the emission of gravitational waves and/or matter lost by centrifugal forces will fight against gravitational collapse. Thus, exceeding a spherically symmetric threshold can be seen as a necessary condition for PBH formation.  For radiation, recent work \cite{yoolate} seems to confirm the existence of a universal threshold related to the volume-averaged compaction function even when the initial curvature perturbations are a-spherical. 

PBHs might also be formed in a variety of other scenarios (see e.g.\cite{others}) where the collapsing fluid equation of state is not that of radiation and perturbations are not necessarily generated during inflation. Thus, in this work we revisit the problem of spherically symmetric black hole formation in a perfect fluid with $p=w \rho$ and  $w \in (0,1]$, with the aim of seeing if the $w=1/3$ analytical results for the threshold \cite{RGE} can be generalized. Note that we only consider the case in which the available time for PBH formation is infinite.  To incorporate these results in the cosmological context, one must also require PBH formation in finite time, and this may make the threshold time-dependent.  This is particularly true in the limiting case of dust ($w\to 0$), where all over-dense perturbations will eventually collapse, although the time to collapse will depend on the value of the compaction function and its shape. On the other hand, the time for PBH formation is known to decrease as $w$ increases, and for $w\gtrsim 1/3$ the time dependence of the threshold is very weak \cite{meilia}.  

Our paper is organised as follows: Sections~\ref{sec:sharp} and~\ref{sec:numerics} describe the initial conditions and the numerical technique we use to simulate BH formation.  Convergence tests are described in an Appendix.  Section~\ref{sec:heuristics} provides heuristic arguments for the range of $w$ over which universality might hold, and the appropriate variables in which this universality is most obviously manifest.  Section~\ref{sec:analytics} provides an analytic formula for the dependence of the critical threshold for BH formation on $w$ and the profile shape.  Sections~\ref{sec:profiles} and~\ref{sec:tests} demonstrate its accuracy using a variety of profile shapes.  Section~\ref{sec:compare} compares our results with previous work for the few special cases where this is possible, and a final section summarizes our findings.
  
\section{Initial conditions for black hole formation}\label{sec:sharp}

We use the Misner-Sharp equations \cite{misnersharp} to simulate the gravitational collapse of cosmological perturbations in spherical symmetry within a Friedman-Robertsnon-Walker (FRW) background. We consider a perfect fluid, $p =w \rho$, with energy momentum tensor $T^{\mu \nu} = \rho(w+1)u^{\mu}u^{\nu}+w \rho g^{\mu\nu}$ and the following metric:
\begin{equation}
\label{metricsharp}
ds^2 = -A(r,t)^2 dt^2+B(r,t)^2 dr^2 + R(r,t)^2 d\Omega^2,
\end{equation}
where $d\Omega^{2} = d\theta^2+\sin^2(\theta) d\phi^2$ is the line element of a 2-sphere and $R(r,t)$ is the areal radius. The components of the four velocity $u^{\mu}$ (which are equal to the unit normal vector orthogonal to the hyperspace at cosmic time $t$ $u^{\mu}=n^{\mu}$), are given by $u^{t}=1/A$ and $u^{i}=0$ for $i=r,\theta,\phi$. 

The Misner-Sharp equations, written in a form that is convenient for numerical simulations (and with $G_{N}=1$), are \cite{albert_paper}: 
\begin{align}
\label{eq:msequations1}
\dot{U} &= -A\left[\frac{w}{1+w}\frac{\Gamma^2}{\rho}\frac{\rho'}{R'} + \frac{M}{R^{2}}+4\pi R w \rho \right], \nonumber\\
\dot{R} &= A U, \\
\dot{\rho} &= -A \rho (1+w) \left(2\frac{U}{R}+\frac{U'}{R'}\right), \nonumber\\
\dot{M} &= -4\pi A w \rho U R^{2}\ ,\nonumber \\
M' &= 4 \pi  \rho R^{2} R', \nonumber
\end{align}
where $(\dot{})$ and $(')$ represent time and radial derivatives respectively. Here $U$ is the radial component of the four-velocity associated to the Eulerian frame and $M$ is the Misner-Sharp mass (which includes contributions from the kinetic energy and gravitational potential energies) introduced as:
\begin{equation}\label{massdef}
M(r,t) \equiv \int_{0}^{R} 4\pi R^{2} \rho \, \left(\frac{\partial R}{\partial r}\right) dr\, ,
\end{equation}
which is related to $\Gamma$, $U$ and $R$ though the constraint:
\begin{equation}
\Gamma = \sqrt{1+U^2-\frac{2 M}{R}}.
\end{equation}

The boundary conditions to this system of differential equations are $R(r=0,t)=0$, leading to $U(r=0,t)=0$ and $M(r=0,t)=0$. Moreover, by spherical symmetry and to ensure regularity of the metric \eqref{metricsharp} at $r=0$, we have $D_{r} \rho (r=0,t)=0$. Finally, in this work we shall only consider type I collapses where $D_r R>0$, as type II are in some sense always over-threshold \cite{kopp}. Because we have a finite grid of size $r_f$, the condition of an asymptotically FRW is imposed by fixing $\rho'(r=r_f,t)=0$.

The lapse function $A(r,t)$ can be solved analytically. Imposing $A(r_f,t) = 1$, to match with the asymptotic FRW spacetime, we have 
\begin{equation}
A(r,t) = \left(\frac{\rho_{b}(t)}{\rho(r,t)}\right)^{\frac{w}{w+1}},
\end{equation}
where $\rho_{b}(t) = \rho_{0}(t_{0}/t)^{2}$ is the energy density of the FRW background and $\rho_{0}=3 H_{0}^{2}/8\pi$.

In addition, to set up the initial conditions for Black Hole (BH) formation, the metric \eqref{metricsharp} at superhorizon scales can be approximated, at leading order in gradient expansion, by \cite{sasaki}:
\begin{equation}
\label{frwmetric}
ds^2 = -dt^2 + a^2(t) \left[\frac{dr^2}{1-K(r) r^2}+r^2 d\Omega^2 \right].
\end{equation}
The cosmological perturbation is encoded in the initial curvature $K(r)$. At leading order in gradient expansion and at super-horizon scales, the product $K(r)r^{2}$ is proportional to the compaction function 
\begin{equation}
 \com(r) \simeq \frac{2 \left[M(r,t)-M_{b}(r,t)\right]}{R(r,t)},
\label{compactionfunction}
\end{equation}
which represents a measure of the mass excess inside a given volume parameterized by $r$ \cite{sasaki}, via the relation
\be
 \com(r)=f(w) K(r) r^{2},
\ee
where
\be
 \label{eq:fw}
 f(w)= \frac{3(1+w)}{(5+3w)}\ .
 \ee
We use $r_m$ to denote the scale on which $\com(r)$ is a maximum.  The value $\com(r_{\rm m})$ on this scale is used as a criterion for PBH formation \cite{refrencia-extra-jaume,sasaki}.  The maximum possible value of $\com(r_{\rm m})$ is $\delta_{c,\rm max}=f(w)$.  This is why $f(w)$ appears explicitly in the expression above.

Specifying the initial conditions corresponds to choosing a particular curvature profile $K(r)$, after which the compaction function $\com(r)$ evolves non-linearly. Whenever $\com(r_{\rm m})>\delta_c(w,{\rm profile})$, the gravitational compression wins against pressure gradients and the expansion of the background universe. This leads inexorably to the formation of a black hole after the first apparent horizon is formed.  Typically, this happens whenever the maximum of the compaction function is of order unity (for a more formal discussion see \cite{muscoelis}, but recall that, in any case, this value cannot exceed $\delta_{c,\rm max}=f(w)$.)

In what follows, we will refer to $\delta_c(w,{\rm profile})$ as the {\it threshold}.  We are particularly interested in quantifying the dependence on $w$ and checking if the dependence on profile shape can be included simply, as it is for $w=1/3$.

\section{Numerical technique}\label{sec:numerics}

In order to numerically solve the system \eqref{eq:msequations1}, we have used the publicly available code based on pseudo-spectral methods \cite{albert_paper}. The time integration of the differential equations is performed with a fourth-order explicit Runge-Kutta method, while the Chebyshev collocation method is used to discretise the grid and evaluate the spatial derivatives \cite{spectrallloyd}.  In this method, the spatial domain is discretised in a Chebyshev grid, whose nodes are given at $x_{k}=\cos(k\pi /N_{\rm cheb})$, where $k=0,1,..,N_{\rm cheb}$ and $N_{\rm cheb}$ is the number of points on the grid. To compute the spatial derivatives at the Chebyshev points we use the Chebyshev differentiation matrix $D$. See \cite{albert_paper} for details.

Pressure gradients increase with increasing $w$ implying the necessity of also increasing the numerical accuracy. Therefore, for $w>1/3$, we have improved the technique developed in \cite{albert_paper} by using a composite Chebyshev grid: we split the full domain in several Chebyshev grids that differ in terms of the necessary density of points to reach the desired accuracy. More technically, our domain is divided into $M$ subdomains given by $\Omega_{l} =[r_{l},r_{l+1}]$ with $l=0,1...,M$. Since the Chebyshev nodes are defined in $[-1,1]$, we also perform a mapping between the spectral and the physical domain for each Chebyshev grid. In particular, we use a linear mapping for each subdomain defined as:
\begin{equation}
 \tilde{x}_{k,l} = \frac{r_{l+1}+r_{l}}{2}+\frac{r_{l+1}-r_{l}}{2}x_{k,l} ,
\end{equation}
where $\tilde{x}_{k,l}$ are the new Chebyshev points re-scaled to the subdomain $\Omega_{l}$. In the same way, the Chebyshev differentiation matrix is re-scaled using the chain rule:
\begin{equation}
 \tilde{D_{l}} = \frac{2}{r_{l+1}-r_{l}} D_{l}.
\end{equation}
Each subdomain is independently evolved in time with the Runge-Kutta 4 methods. The spatial derivative in each subdomain is computed by the associated Chebyshev differentiation matrix $\tilde{D_{l}}$.

In order to evolve across different $\Omega_l$s we need to impose boundary conditions. For this, we have followed the approach of \cite{bh_spectral}. At the boundaries between subdomains, the time derivative of each field is computed. Then, the incoming fields derivative is replaced by the time derivatives of the outgoing fields from the neighboring domain. Following an analysis of the characteristics like the one performed in \cite{bh_spectral}, we have checked that all the fields are incoming except for the density field, which is directed outwards. Appendix~\ref{sec:testsims} describes convergence tests which give us confidence in our numerical simulations.

\section{Analytical threshold: heuristic arguments}\label{sec:heuristics}

As explained in \cite{RGE}, to a very good approximation, the threshold for the $w=1/3$ case only depends upon the curvature of the compaction function at its maximum, under the assumption of a central over-dense peak in the density distribution. Here we give a slightly different heuristic argument for why this is so and also draw some conclusions about the cases $w\neq 1/3$. 

\subsection{Shape approximation}
Following \cite{harada}, we first crudely model a sharply peaked initial density distribution as a homogeneous core (a closed universe) surrounded by a thin under-dense shell between it and the external expanding universe.

The speed of propagation in a closed FRW universe is equation-of-state dependent:
\be\label{velo}
 v=\frac{\sqrt{w}}{1+3 w}\ .
\ee 
This speed has a maximum at $w=1/3$, from which it falls relatively steeply for $w<1/3$ and less steeply for $w>1/3$.  For radiation ($w=1/3$), only a very small portion around the maximum of the gravitational potential (which is typically at the border of the core) will contribute to the collapse.  All other surrounding fluid-elements will manage to escape the gravitational attraction. However, if the equation of state differs from $w=1/3$, a larger portion of the fluid will participate in the collapse.  Hence, as $w$ becomes increasingly different from $1/3$, we may expect the threshold to depend more and more on the full shape of the compaction function. Moreover, this dependence will be asymmetric:  we expect a stronger dependence for $w<1/3$ than $w>1/3$. This is indeed what we are going to show numerically.

\begin{figure}[b]
\includegraphics[width=0.6\linewidth]{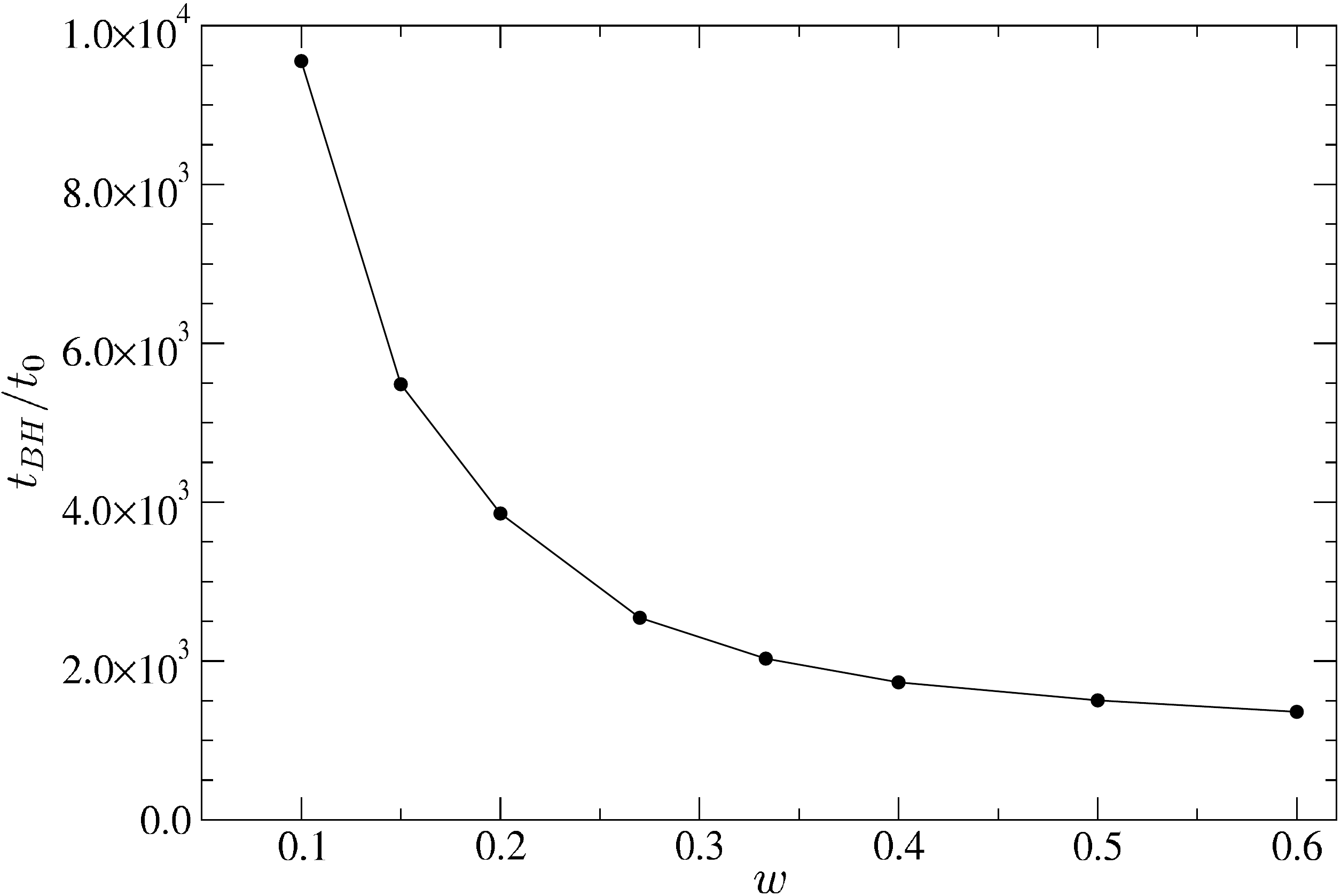} 
\caption{Dependence on $w$ of the time for a perturbation to collapse and form an apparent horizon. For this example the initial perturbation (at $t=t_0$) is given by Eq.\eqref{basis_pol} with $q=1$ and $\delta = \delta_{c}+10^{-2}$.}
\label{fig:time_colapse}
\end{figure}

If the escape velocity were the only ingredient, the point of maximal velocity would also correspond to the maximal threshold, as reported in \cite{harada}. This, however, does not make sense \cite{domenech}: the approximation of \cite{harada} misses the fact that if the density is inhomogeneous, then this generates gradient pressures that are larger if $w$ is large. These resist the collapse, so we might expect the threshold to increase with $w$.  However, even this is not the full story. Pressure gradients are also a form of gravitational energy so, while they initially work against the collapse, once the collapse is triggered, they mostly favor it. The net result is a smaller formation time for a larger $w$, as can be seen in Fig.\ref{fig:time_colapse}.

To summarize:  Our heuristic arguments suggest that the methodology of \cite{RGE} for finding a universal threshold might also be useful for $w>1/3$ but it is likely to fail for $w<1/3$.  

\subsection{Use of average compaction function}

At super-horizon scales, the perturbations at threshold are very well approximated by their Newtonian counterpart. Because the space and time dependence of the perturbation decouples, one has that
\be\label{phi}
 \nabla^2 \Phi = 8\pi \bar\rho\ ,
\ee 
where $\bar\rho(r)\equiv(a H)^2\frac{\delta\rho(r,t)}{\rho_b}$,
$\Phi$ is the Newtonian potential and $\nabla^2$ is the Euclidean Laplacian.
Eq~\eqref{phi} is solved by
\be
 \Phi(r)=8\pi\int_{0}^r \frac{dx}{x^2} \int_0^{x} dy\,y^2\,\bar\rho\ .
\ee 
In this limit the compaction function is 
\be
 \com(r)=\frac{3}{r}\int_0^r dy y^2 \bar\rho\ ,
\ee
and thus
\be
 \Phi(r)=\frac{8\pi}{3}\int_0^r \frac{\com(x)}{x}dx\ .
\ee
Now suppose only the potential difference around $r_m$ is important for the gravitational collapse. Then we can consider the difference $\Phi(r_m)-\Phi(r_0)$ where $r_0\equiv r_m(1-\alpha)$. Assuming this region is weakly dependent upon the profile chosen, once the equation of state is fixed, we can approximate $\alpha\simeq\alpha(w)<1$. Then, 
\be\nonumber
 \Phi(r_m)-\Phi(r_0)=\frac{8\pi}{3 V_{\alpha}}\int_{r_m(1-\alpha)}^{r_m}
                        x^2\com(x)\frac{V_{\alpha}}{x^3}dx\ ,
\ee
where $V_{\alpha}$ is the volume in the shell of internal radius $r_m(1-\alpha)$ and external radius $r_m$.
Since $\alpha< 1$, we have 
\be
 \Phi(r_m)\simeq\alpha\frac{8\pi}{3}\bar\com+{\cal O}(\alpha^2)\ .
\ee
This shows that if the gravitational collapse only depends on the potential difference around the maximum of the compaction function, then the threshold will mainly depend on the volume averaged compaction function, and not on the other details of its profile.  Because of this, one could equivalently study the dual problem of a top-hat compaction function with height equal to the average of the original compaction function. This is precisely what we did in \cite{RGE}\footnote{In \cite{RGE} we showed that setting $\alpha=1$ works well. Here we show that allowing $\alpha<1$ leads to a better approximation.}.

\section{Analytic formula for the threshold}\label{sec:analytics}

In this section, we suppose that the equation of state of the fluid is such that it allows us to expand the compaction function around its maximum ($r=r_m$). Then, as in \cite{RGE}, to a very good approximation the threshold only depends on
\be\label{q}
 q\equiv -\frac{r_m^2\,\com''(r_m)}{4\,\com(r_m)} \ ,
\ee
which is a dimensionless measure of the curvature of $\com(r)$ at its maximum. 

To proceed, we define a ``basis'' (or fiducial set of curvature profiles) such that, by varying $q$, this set covers the whole range of interesting thresholds and shapes with $q\in (0, \infty)$ while also being regular at $r=0$ and having $\rho'(r=0,t)=0$. In \cite{RGE}, this basis was given in terms of the exponential functions used previously by \cite{ilia}.  However, because the boundary conditions at the origin are violated for $q<0.5$, we instead consider the basis
\begin{equation}
\label{basis_pol}
 K_{\rm b}(r) =\frac{\com(r_m)}{f(w)r_m^2}\frac{1 + 1/q}{1+\frac{1}{q}\left(\frac{r}{r_{m}}\right)^{2(q+1)}}\ .
\end{equation}
This fiducial set satisfies the appropriate boundary and regularity conditions for any $q>0$.

We then define 
\begin{equation}
  \com_{\rm b}(r) = f(w) r^2 K_{\rm b}(r)\ .
\end{equation} 
The critical compaction function, averaged within a spherical shell extending from radius $[1-\alpha(w)]\,r_m$ to $r_m$, is defined to be 
\be
\label{Cbar}
 \bar\com_{\rm c}(w,{\rm profile})\equiv
 \frac{3}{ r_m^3 V[\alpha(w)]}\int_{r_m[1-\alpha(w)]}^{ r_m} \com_c(r) r^2 dr\ ,
\ee
where $V[\alpha(w)]=\alpha(w)\,[3+(\alpha(w)-3)\alpha(w)]$ and $\com_c(r)=\com(r)\Big|_{\com(r_m)=\delta_c}$.

Inserting Eq.\eqref{basis_pol} in Eq.\eqref{Cbar} yields 
\be
 \label{average_formula}
 \bar\com_{\rm c}(w, {\rm basis})=  \delta_c(w,q)\, g(q,w)\,
   \left[-F_1(q)+(1-\alpha)^{3-2q}F_2(q,\alpha) \right]  ,
\ee
with 
\begin{equation}
 g(q,w) = \frac{3(1+q)}{\alpha (2q-3)\left[3+\alpha(\alpha-3)\right]} ,
\end{equation}
\begin{equation}
 F_1(q) = {}_2F_1\left[ 1,1-\frac{5}{2(1+q)}, 2-\frac{5}{2(1+q)},-q \right] ,
\end{equation}
and
\begin{equation}
 F_2(q,w) = {}_2F_1\left[1,1-\frac{5}{2(1+q)},2-\frac{5}{2(1+q)},-q(1-\alpha)^{-2(1+q)}\right] ,
\end{equation}
where ${}_2F_1$ is the hypergeometric function.

Notice that if 
\be\label{assumption}
 \bar\com_{\rm c}(w,{\rm profile})\simeq \bar\com_{\rm c}(w)\ ,
\ee
i.e. if the dependence of the averaged critical compaction function on profile shape is weak enough to be ignored, then one can simply rearrange Eq.~\eqref{average_formula} to obtain an analytic expression for the critical threshold value:  
\begin{equation}
 \label{threshold-anal}
 \delta_c^A(w,q) = \frac{\bar\com_{\rm c}(w)}{ g(q,w)}\, \frac{1}{\left[-F_1(q)+(1-\alpha)^{3-2q}F_2(q,\alpha) \right]}\ .
\end{equation}
Once $\alpha(w)$ has been specified, Eq.~\eqref{threshold-anal} represents our generalization of \cite{RGE} to $w\ge 1/3$. 

In \cite{RGE}, where $w=1/3$, $\alpha$ was a constant set equal to $1$ and hence $\bar\com_{\rm c}$ equaled the volume average within the sphere of radius $r_m$. Here, we allow $\alpha$ to depend on $w$ but we still assume its dependence on $q$ to be negligible. As we shall see, this assumption is good enough only for $w\gtrsim 1/3$, as we suggested in the previous section. In particular, we shall find that even for the case $w=1/3$, the optimal $\alpha$ is smaller than $1$. In this sense, the current analysis not only generalizes the work of \cite{RGE} to $w\ne 1/3$, it also enhances the precision of the $w=1/3$ case.

\subsection{The appropriate volume over which to average}

\begin{figure}[t]
	\centering
	\includegraphics[width=0.45\linewidth]{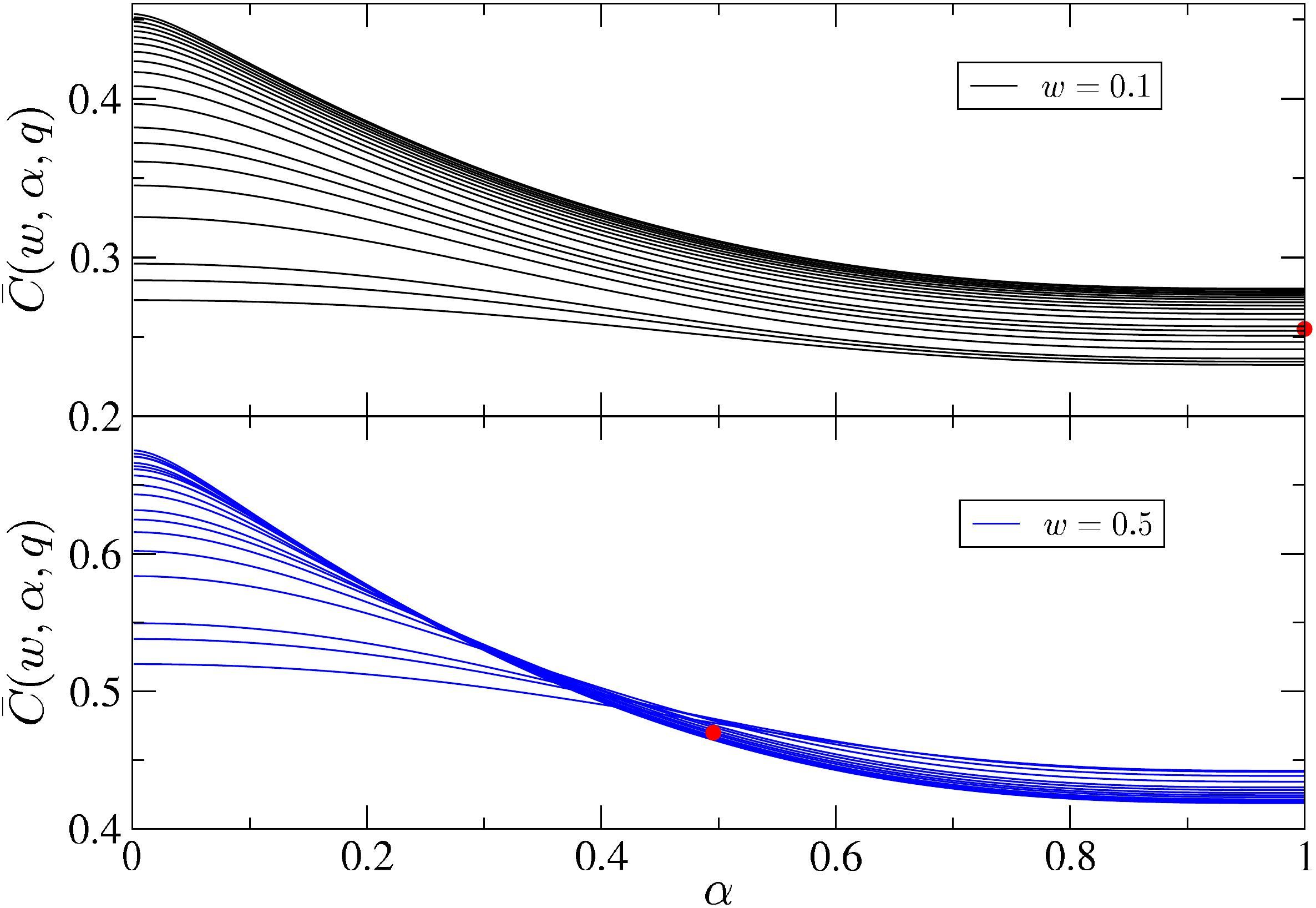} 
	\includegraphics[width=0.45\linewidth]{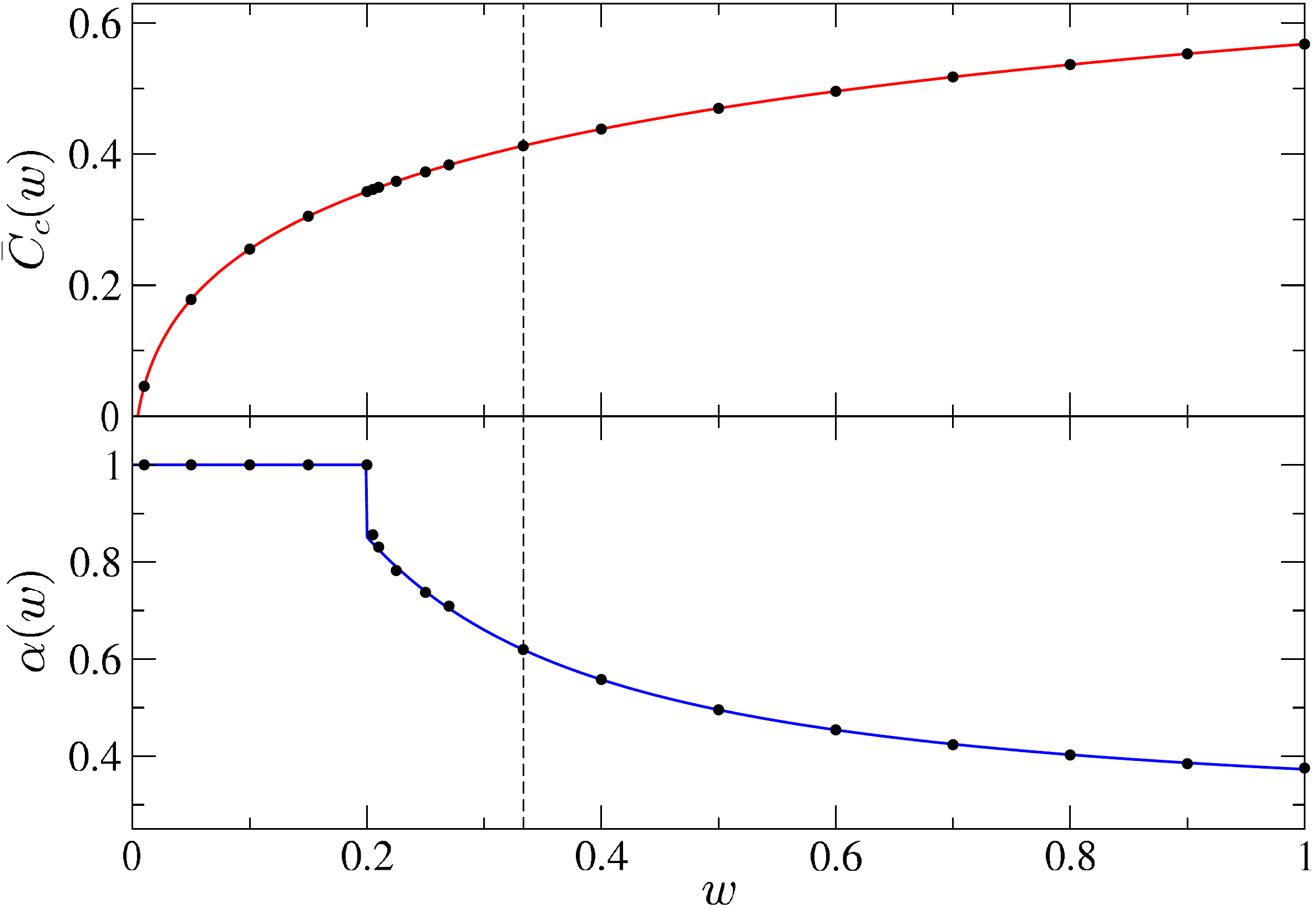} 
	\caption{Left:  Dependence of $\bar\com$ on the volume within which it is averaged, for two choices of $w$ (top and bottom panels) and a variety of basis shapes (curves show different $q$'s) for each $w$. Red circle in each panel shows the pair $(\alpha,\bar\com)$, Eqs.\eqref{eq:average_c} and~\eqref{eq:alpha} respectively, which return the best estimates of $\delta_c$ when inserted in our universal threshold formula (Eq.\ref{threshold-anal}).  Right:  Symbols in top and bottom panels show $\bar{\com}(w)$ and $\alpha(w)$ for profiles given by Eq.\eqref{basis_pol}; curves show Eqs.\eqref{eq:average_c} and~\eqref{eq:alpha}. Vertical dashed line is at $w=1/3$.}
	\label{fig:fitting_basis}
\end{figure}

We determine $\alpha(w)$ as follows: 
Consider a family of profiles parameterized only by $\com(r_m)$ and $q$, such as those given by Eq.\eqref{basis_pol}.  We evolve each profile using the code described in Section~\ref{sec:numerics}, and hence determine the threshold $\delta^N_c(w,q)$.  We then perform the volume integral for various $\alpha$ to find the corresponding $\bar\com_c(w,\alpha,q)$.  
The left panel of Fig.(\ref{fig:fitting_basis}) illustrates:  the top and bottom panels show results for different $w$; the different curves in each panel show how $\bar \com_c(w,\alpha,q)$ varies with $\alpha$ as $q$ is increased  in steps of $ \approx 1$, when the profile shape is given by Eq.\eqref{basis_pol}. The top left panel shows that $\bar\com_c(w,\alpha,q)$ can vary by tens of percent with $q$ when $w=0.1$.  However, the bottom left panel shows that this variation is much smaller when $w=0.5$; at $\alpha\approx 0.5$, $\bar\com_c(w,\alpha,q)$ varies by less than 5\% for the entire range of $q$ we have considered.  This is consistent with the heuristics of the previous section, which argued that details of the profile shape should matter much more at small $w$.  

Since the dependence on $q$ is weak, we have parametrized the remaining dependence on $w$ (comparison of the top and bottom panels shows that $\bar\com_c$ tends to be larger for larger $w$) as follows:  To minimize the error associated with using $q$-{\it independent} $\alpha$ and $\bar\com_c$ values in Eq.\eqref{threshold-anal}, we first chose the value of $\alpha(w)$ corresponding to the point where the flux of $\bar\com_c(w,\alpha,q)$ (e.g., in the bottom left panel) is densest. Once $\alpha(w)$ is given, the $q$-independent $\com_c(w)$ is chosen to minimize the difference between its value and the numerical $q$-dependent ones. The red circle at $\alpha\approx 0.5$ in the bottom left panel of Fig.(\ref{fig:fitting_basis}) shows the result of this double minimization for $w=0.5$.  The red circle in the top left panel is at $\alpha=1$.  We discuss the significance of this difference shortly.

The symbols in the right hand panels of Fig.(\ref{fig:fitting_basis}) show $\alpha(w)$ and $\com_c(w)$ resulting from following this procedure for the basis profiles (Eq.\ref{basis_pol}).  They show that $\bar\com_c$ decreases monotonically with $w$; the limit $\bar\com_c(w \rightarrow 0) = 0$ reflects the fact that $\delta_{c}(w \rightarrow 0) = 0$.  Instead, $\alpha$ increases as $w$ decreases reaching its maximal value, unity, for $w \lesssim 0.2$.  Larger values of $\alpha$ indicate that the threshold is sensitive to the whole profile shape rather than just $q$ (which describes the profile shape at $\alpha\to 0$).  Thus, the increase of $\alpha$ as $w$ decreases, and the fact that $\alpha\to 1$ for $w<1/3$, are in qualitative agreement with the discussion of the previous section.

The trends shown in the right hand panels are well described by 
\begin{align}
 \label{eq:average_c}
 \bar\com_{\rm c}(w) &= a + b \, {\rm Arctan}(c\, w^d)  \\
 \label{eq:alpha}
 \alpha(w) &= e + f \, {\rm Arctan}(g\, w^h) ,
\end{align}
with $a = -0.140381$, $b = 0.79538$, $c=1.23593$, $d= 0.357491$, $e = 2.00804$, $f = -1.10936$, $g = 10.2801$ and $h=1.113$. 
Inserting Eqs.\eqref{eq:average_c} and~\eqref{eq:alpha} in Eq.\eqref{threshold-anal} yields an analytic expression for $\delta_{c}(q,w)$. 
To connect with \cite{RGE}, note that when $w=1/3$ we have $\alpha\sim 0.6$ and $\bar{\com}_c\sim 0.4$.  This value of $\bar{\com}_c$ is similar to that obtained by \cite{RGE} who explicitly set $\alpha=1$. Eq. \eqref{eq:average_c} is then our generalization of the \cite{RGE} analysis to $w>1/3$.

Having established that our methodology works for profiles of the form Eq.\eqref{basis_pol}, the next section tests its accuracy and generality.  However, before moving on, we note that there is a technical issue with the basis Eq.\eqref{basis_pol}.  As $q\to 0$, $\com_{\rm b}(r)$ becomes nearly constant over an ever wider range of scales. Because our simulation uses only a finite number of grid points, the non-zero constant compaction function at the grid  ``infinity'' -- i.e. on the scale of the box -- results in a fictitious conical singularity which violates the boundary condition of a flat FRW.  For our simulations, this occurs when $q<0.1$.  In addition, for $q\gg 1$, $K_{\rm b}$ becomes close to a tophat, and $\com_{\rm b}$ becomes sharply peaked at $r_m$.  This results in  pressure gradients which are difficult to simulate accurately.  For this reason, Eqs.\eqref{eq:average_c} and~\ref{eq:alpha} have really only been calibrated using simulations over the range $q\in \left[0.1,30\right]$.  Of course, this restriction on the range of $q$ is not physical:  in principle smaller $q$ can be simulated simply by using more grid points.  Rather than paying the larger computational price of longer run times as one moves to more and more grid points, in the next sections we check that extrapolating our results to $q<0.1$ agrees with simulations of other profiles which have low $q$ but for which the fictitious singularity at low $q$ does not arise.  We also consider the $q\to\infty$ limit in more detail later.

\section{Choice of profile shape}\label{sec:profiles}
Here we test the approximation that both $\alpha$ and $\bar\com_c$ only depend on $w$. To do so we consider three other families of curvature profiles:
\begin{align}
\label{eq:exp}
K_{1} &=\frac{\com(r_m)}{f(w)r_m^2}e^{\frac{1}{q}\left(1-\left[\frac{r}{r_m}\right]^{2q}\right)} ; \\
\label{eq:lamda}
K_{2} &= \frac{\com(r_m)}{f(w) r_m^2}\,
\left(\frac{r}{r_{m}}\right)^{2\lambda}\,
e^{\frac{(1+\lambda)^{2}}{q}\left(1 - \left(\frac{r}{r_{m}}\right)^{\frac{2q}{1+\lambda}}\right)} ; \\
\label{eq:spectrum}
K_{3} &= \frac{ \com(r_m)}{f(w) r_m^2}\frac{r_m^3}{r^3}\,
\frac{g(n(q),k_p,r)}{g(n(q),k_p,r_m)} ,
\end{align}
where
\begin{align}
  g(n(q),k_p,r) &=\Lambda^{3+n}g_{1}(n(q),k_p,r) +g_{2}(n(q),\Lambda,k_p,r)  ,
  \qquad {\rm with}\nonumber\\
  g_{1}(n(q),k_p,r) &= \left[k_{p}r \left\{ E_{3+n}(-ik_{p}r)+E_{3+n}(ik_{p}r)\right\} + i \left\{-E_{4+n}(ik_{p}r)+E_{4+n}(-ik_{p}r) \right\} \right]\ ,\nonumber\\
  g_{2}(n(q),\Lambda,k_p,r) =& \left[-\Lambda k_{p}r  \left\{ E_{3+n}(-i \Lambda k_{p}r)+E_{3+n}(i \Lambda k_{p}r)\right\} - i \left\{-E_{4+n}(i \Lambda k_{p}r)+E_{4+n}(-i \Lambda k_{p}r) \right\} \right]\ ,\nonumber
\end{align}
and $E_{n}(x) \equiv \int_{1}^{\infty}e^{-x t}\,dt/t^{n}$.
$K_{1}$ and $K_2$ are the centrally and non-centrally peaked families of exponential profiles discussed in \cite{ilia}, while the oscillating profiles $K_{3}$ are more physically related to models of inflation \cite{Atal-Germani}. There, $\Lambda$ is a UV cut-off of the power spectrum and $k_p$ the Fourier mode related to its highest peak. For $n>0$, one may remove the cut-off in $K_3$ ($\Lambda \rightarrow \infty$). In this case, only the term $g_{1}(n(q),k_p,r)$ would contribute to the curvature profile.

\begin{figure}[t]
	\centering
	\includegraphics[width=0.45\linewidth]{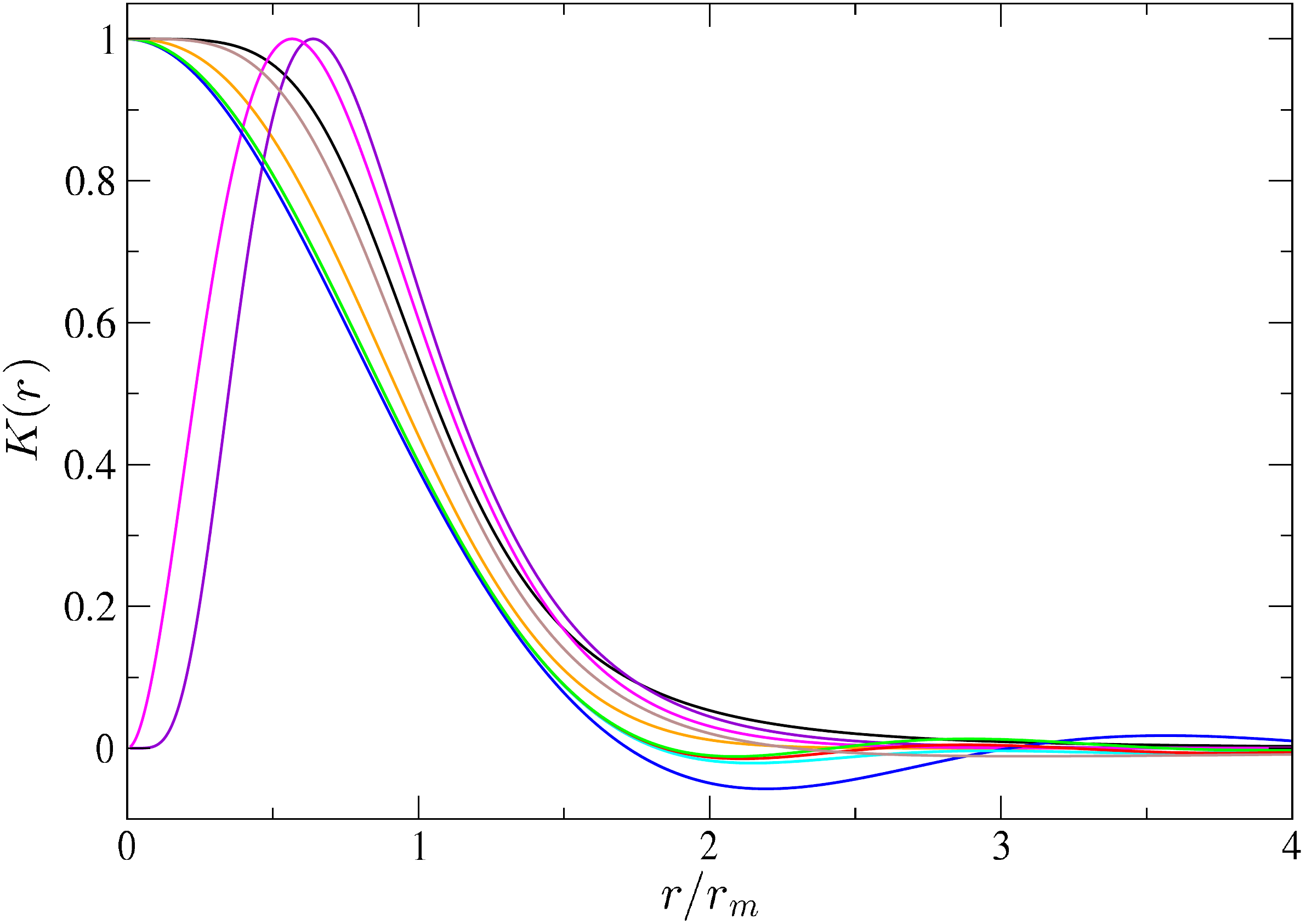} 
	\includegraphics[width=0.45\linewidth]{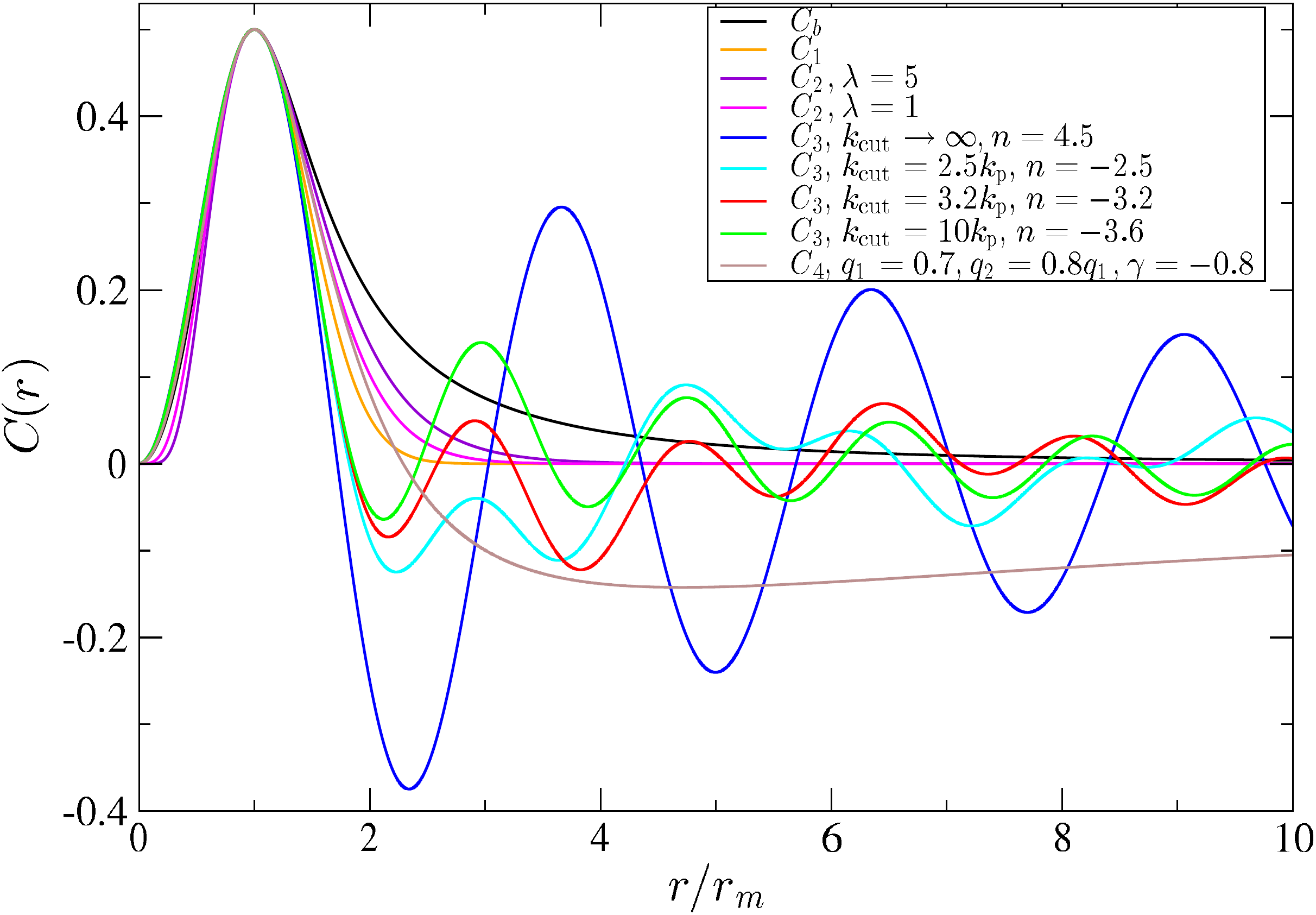} 
	\caption{Illustrative $K(r)$ with the peak normalized to $1$ (left) and corresponding $\com(r)$ (right) profiles associated with Eqs.\eqref{eq:exp}--\eqref{eq:spectrum} with parameters chosen to all have $q=1.22$ at $r_m$ and normalized to $\delta=0.5$.}
	\label{fig:C_tested}
\end{figure}

In the next section we also consider profiles of the form 
\begin{equation}
\label{eq:LC}
 K_{4} =\frac{\com(r_m)}{f(w)r_m^2}\frac{r_m^2}{r^2} \frac{C_{\rm LC}(r)}{C_{\rm LC}(r_m)}, \qquad {\rm with}\qquad 
 C_{\rm LC}(r) =\frac{1 + 1/q_1}{1+\frac{1}{q_1}\left(\frac{r}{r_{m,1}}\right)^{2(q_1+1)}} + \gamma \, \frac{1 + 1/q_2}{1+\frac{1}{q_2}\left(\frac{r}{r_{m,2}}\right)^{2(q_2+1)}} .
\end{equation}
These $K_4$ are a linear combination of two of our basis $K_{\rm b}$ profiles (Eq.\ref{basis_pol}), each having different $q$ and $r_m$.  Our main interest in this family is that the resulting $q<0.1$ is well-behaved without having to use extremely large grids.

\begin{figure}[t]
 \centering
 \includegraphics[width=0.6\linewidth]{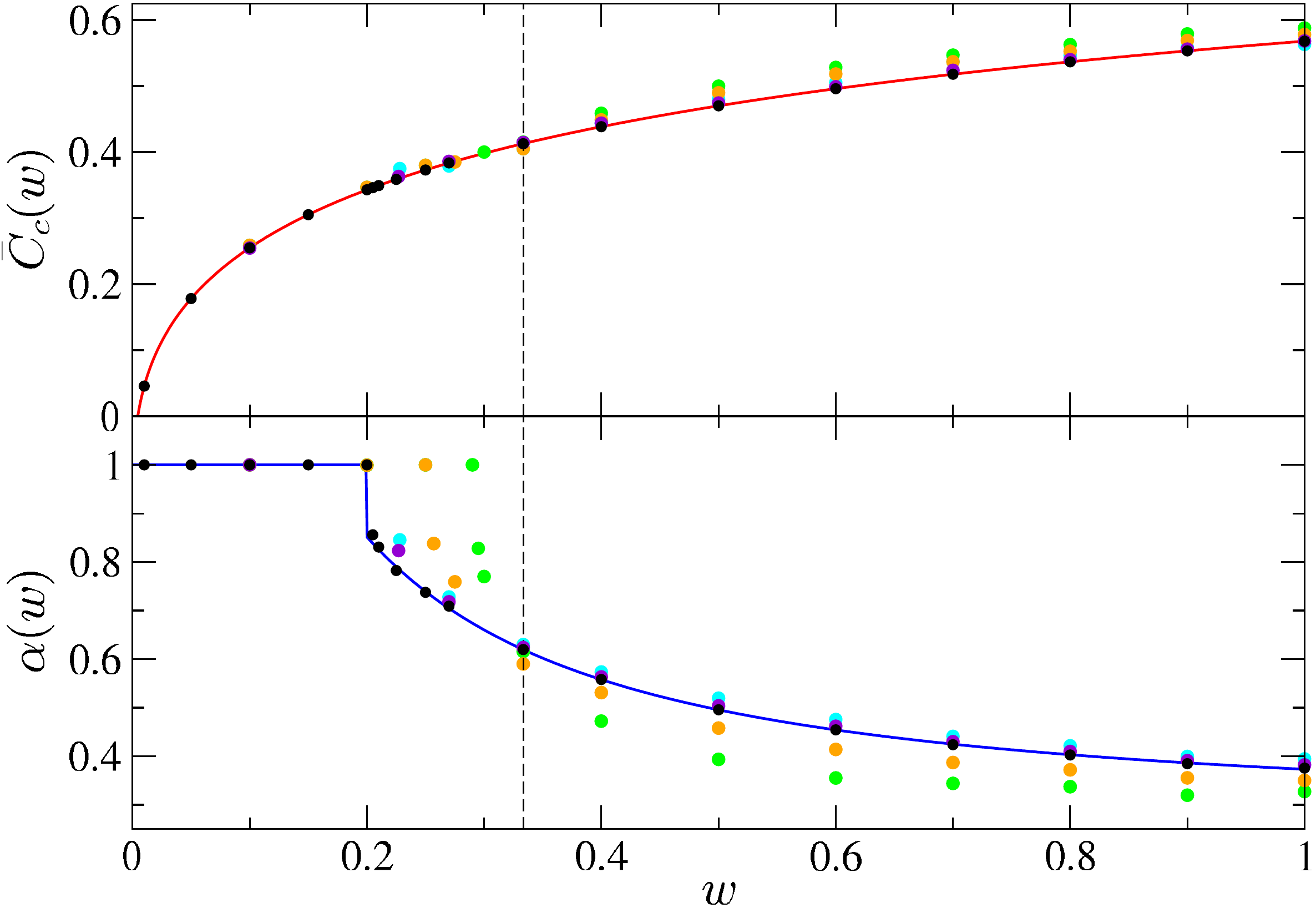} 
 \caption{Same as Fig.\ref{fig:fitting_basis} except that green circles are obtained from simulations in which the initial profiles were described by Eq.\eqref{eq:exp}, orange circles are for Eq.\eqref{eq:lamda} with $\lambda=1$, cyan circles show results for Eq.\eqref{eq:lamda} with  $\lambda=2$ and violet circles are for Eq.\eqref{eq:LC}.  Solid curves show Eqs.\eqref{eq:average_c} and~\ref{eq:alpha} which provide an excellent description of our basis set (Eq.\ref{basis_pol}) based simulations.}
 \label{fig:results_fit2}
\end{figure} 

\begin{figure}[t]
 \centering
 \includegraphics[width=0.6\linewidth]{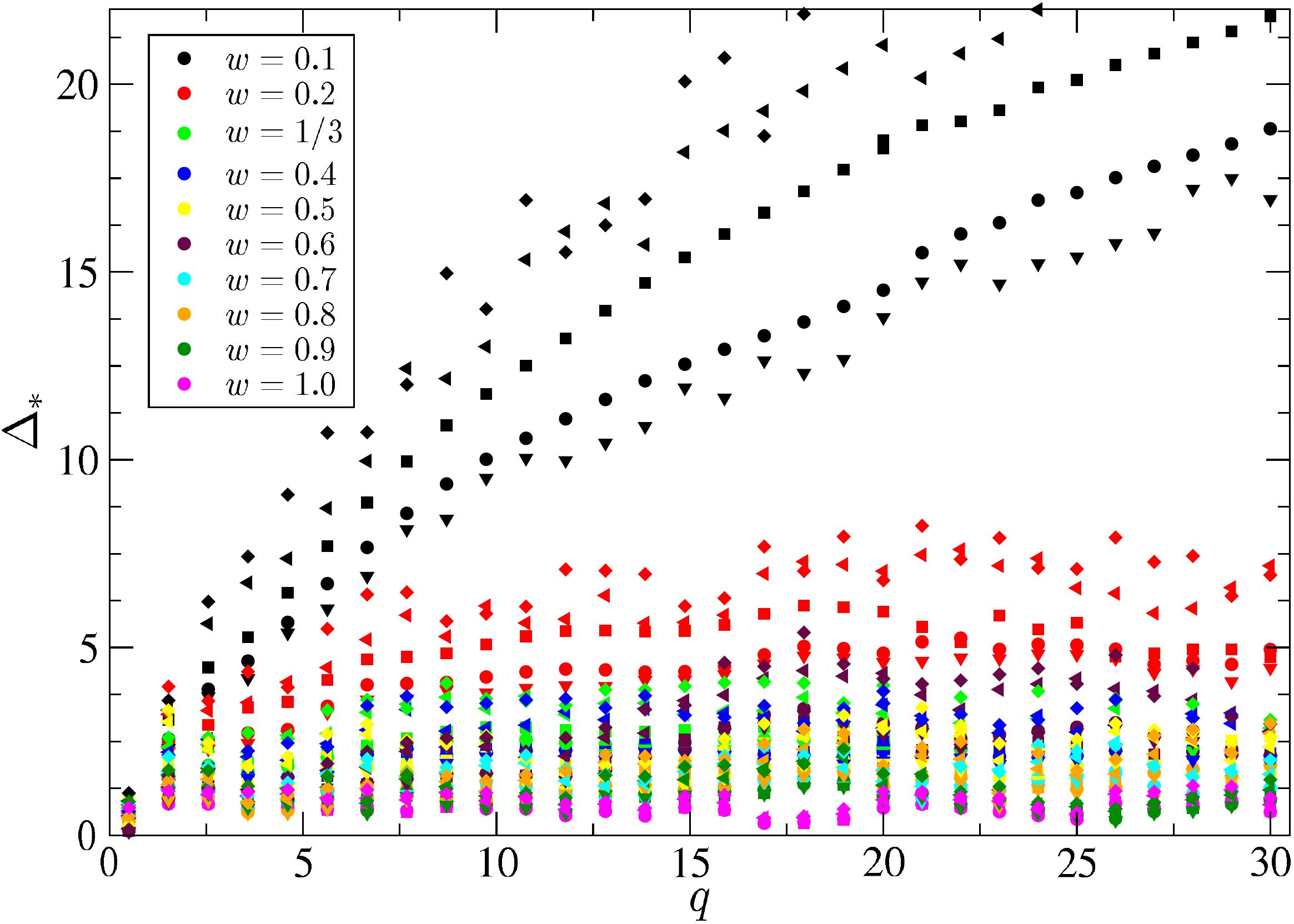} 
 \caption{Relative difference ($\Delta_*$ of Eq.\ref{eq:D*}) between the numerically simulated values $\delta^{N}_{c}$ for the basis Eq.\eqref{basis_pol} and for Eq.\eqref{eq:lamda} with $\lambda=0.5$ (upside down triangles), $\lambda=1$ (solid dots), $\lambda=2$ (squares), $\lambda=5$ (leftward pointing triangles) and $\lambda=10$ (diamonds) for a range of values of $w$ and $q$.}
 \label{fig:dev_numerical_2}
\end{figure} 

Fig.\ref{fig:C_tested} compares a few of these curvature profiles and their associated compaction functions for a variety of parameter choices.  This makes the point that our analysis considers a wide variety of profile shapes.  Then, following the procedure outlined in the previous section, we ran simulations with these other profile shapes, and so obtained the family-dependent $\alpha$s and $\bar\com_c$s.  Finally, we checked if the averaged compaction functions depend mainly on the curvature of $K$ around $r_m$ (i.e. on $q$) or if the full shape between 0 and $r_m$ matters.

Fig. \ref{fig:results_fit2} shows the results.  As we expected, universality -- results which do not depend on the choice of $K$, provided $q$ is fixed -- is most closely achieved when $w=1/3$.  For $w<1/3$, $\alpha$ and $\bar\com_c$ depend strongly on the family of profiles chosen and $\alpha$ quickly saturates to $1$. This is because for small pressure gradients (small $w$) local structure in the initial profile shape matters more.  Therefore, the shape around the peak of $\com$ is no longer the only relevant quantity.  However, notice that for $w>1/3$, the dependence of $\alpha$ and $\bar\com_c$ on choice of parametrization of the initial curvature profile is weak enough to be neglected, as we discuss further below.

To quantify the dependence of $\delta_c$ on choice of $K$ for a given $w$ and $q$, we define
\be
 \Delta_*\equiv 100\,
    \frac{\Big|\delta_c({\rm basis}|q,w)-\delta_c({\rm other\ family}|q,w)\Big|}
         {\delta_c({\rm basis}|q,w)}\ ;
 \label{eq:D*}
\ee 
this is the percent difference between $\delta_c$ returned by the simulations for the fiducial, basis profile and one from another family (having the same $q$ and $w$).  Fig.\ref{fig:dev_numerical_2} shows $\Delta_*$ when the other family is given by Eq.\eqref{eq:lamda}, for a variety of choices of $\lambda$.  For $w<1/3$, $\Delta_*$ clearly depends strongly on both $\lambda$ and $q$.  However, as $w$ increases, $\Delta_*$ decreases and is much less dependent on either $\lambda$ or $q$, with differences down at the one percent level when $w=1$.  This also happens if we replace profiles of the Eq.\eqref{eq:lamda} family with those of Eq.\eqref{eq:exp} or Eq.\eqref{eq:spectrum}.

\begin{figure}[t]
\centering
  \includegraphics[width=0.45\linewidth]{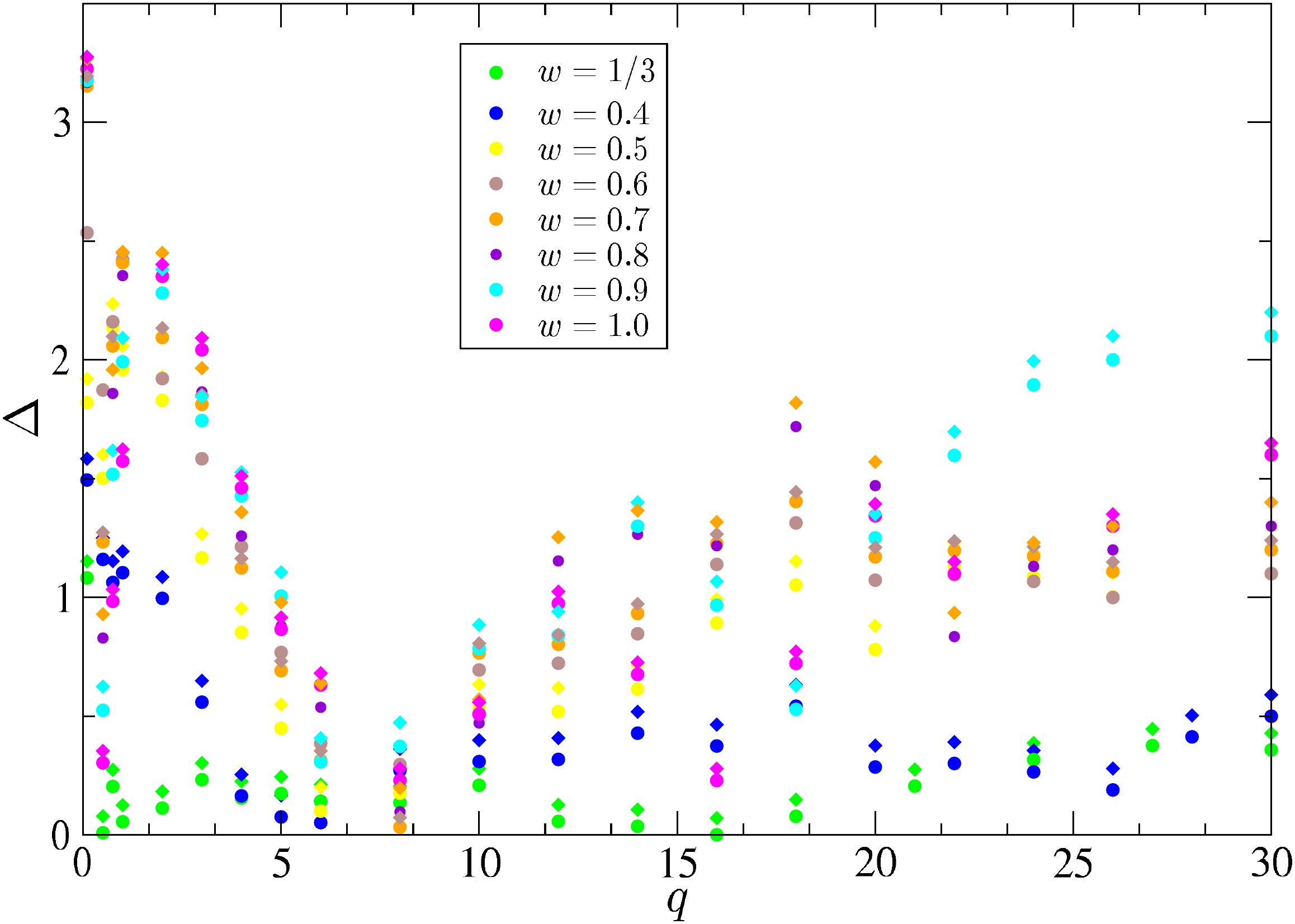} 
  \includegraphics[width=0.45\linewidth]{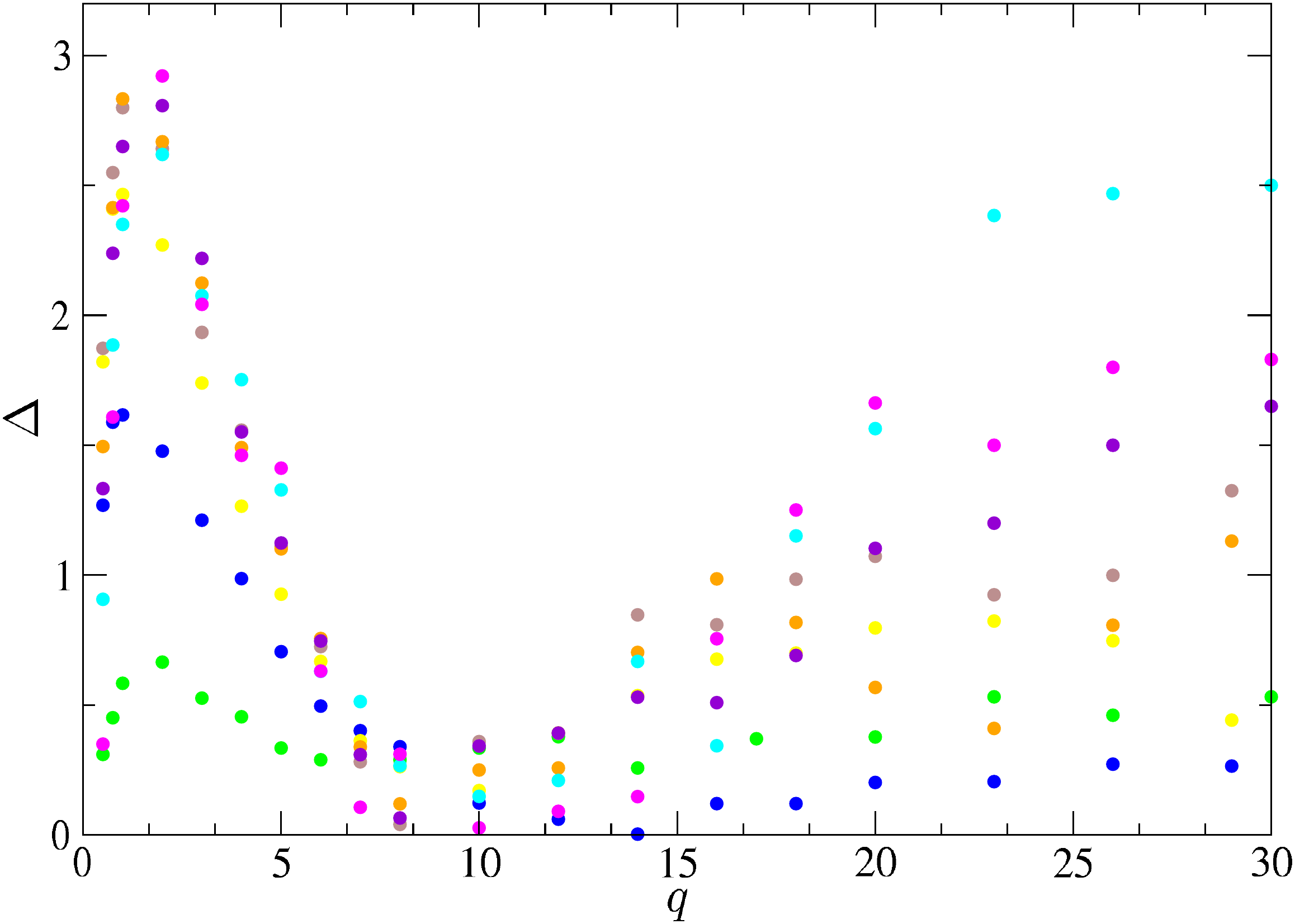} 
  \includegraphics[width=0.45\linewidth]{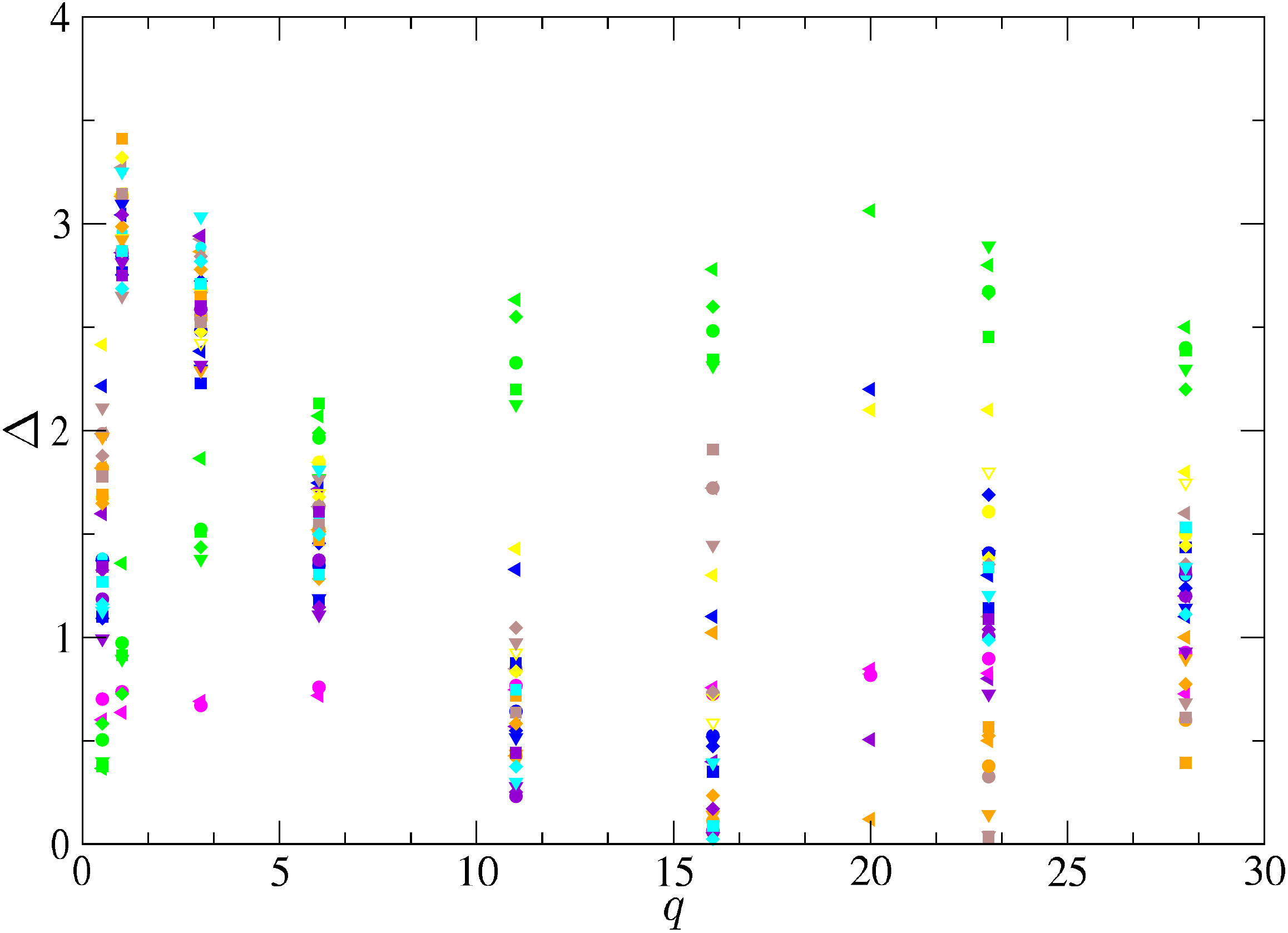} 
\includegraphics[width=0.45\linewidth]{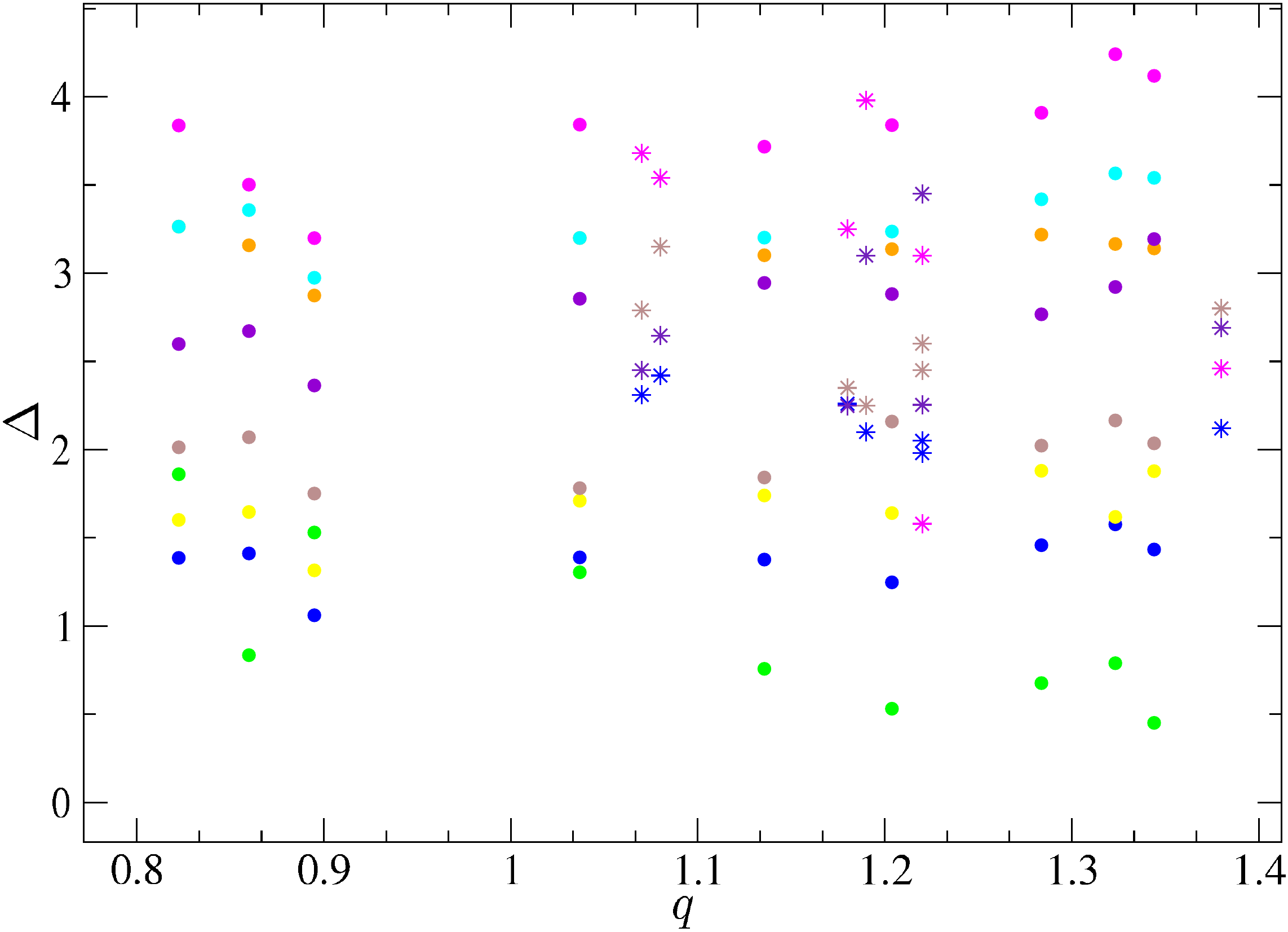} 
\caption{Relative difference $\Delta$ of Eq.\eqref{d} between the analytic values $\delta^{A}_{c}$ \eqref{threshold-anal} and the numerically simulated $\delta^{N}_{c}$ for a range of $w$ and $q$.  Top left panel shows results for the fiducial family of profiles Eq.\ref{basis_pol} (circles) and profiles described by Eq.\ref{eq:LC} (diamonds); top right is when the profile is given by \eqref{eq:exp}; bottom left is for Eq.\eqref{eq:lamda} with $\lambda=0.5$ (upside-down triangles), $\lambda=1$ (solid dots), $\lambda=2$ (triangles pointing left), $\lambda=5$ (squares) and $\lambda=10$ (diamonds); bottom right is for Eq.\eqref{eq:spectrum} with $n\in[0.5,15]$ for $\Lambda\rightarrow \infty$ (solid points) and for $\Lambda \neq \infty$ and $n<0$ (stars).}
\label{fig:dev}
\end{figure}

\section{Numerical versus analytical thresholds for $w\geq 1/3$}\label{sec:tests}
We are now ready to test if our methodology for obtaining an analytic fitting formula for the threshold works, albeit only for $w\geq 1/3$.  To do so, we define 
\be\label{d}
 \Delta\equiv 100\, \frac{\Big|\delta^N_c-\delta_c^A\Big|}{\delta^N_c}\ ,
\ee
where $N$ and $A$ stand for the threshold obtained from the numerical simulation and the corresponding analytic approximation to it given by Eq.\eqref{threshold-anal}.  

The top left panel of Fig.\ref{fig:dev} shows that $\Delta$ of Eq.\eqref{d} is typically less than 6 -- the numerical and analytical thresholds agree at better than the 6\% level --  over the entire range of $w$ and $q$ we have tested. The other panels show the agreement is similarly good for the other families of profiles:  Eqs.\eqref{eq:exp}--\eqref{eq:spectrum}. 
Our results for radiation ($w=1/3$), which make use of the basis Eq.\eqref{basis_pol}, turn out to be slightly more accurate than those of our earlier work \cite{RGE} where the exponential basis, Eq.\eqref{eq:exp}, was used. 

We noted previously that numerical stability and speed make it difficult to estimate $\delta_c$ in simulations with $q\lesssim 0.1$ or $q\gtrsim 30$, due to a conical singularity and large pressure gradients respectively. However, it turns out that the $q\to 0$ and $q\to\infty$ limits are both amenable to further analysis as we now discuss.  In addition to pedagogy, understanding the full range of $q$ is important because, in some models of PBH abundances (e.g. \cite{ravi-cri}), larger $q$ contribute at later times, so the full range of $q$ matters for PBH abundances.  

\subsection{The sharply peaked limit:  $q\to\infty$}
\begin{figure}
  \centering
  \includegraphics[width=0.45\linewidth]{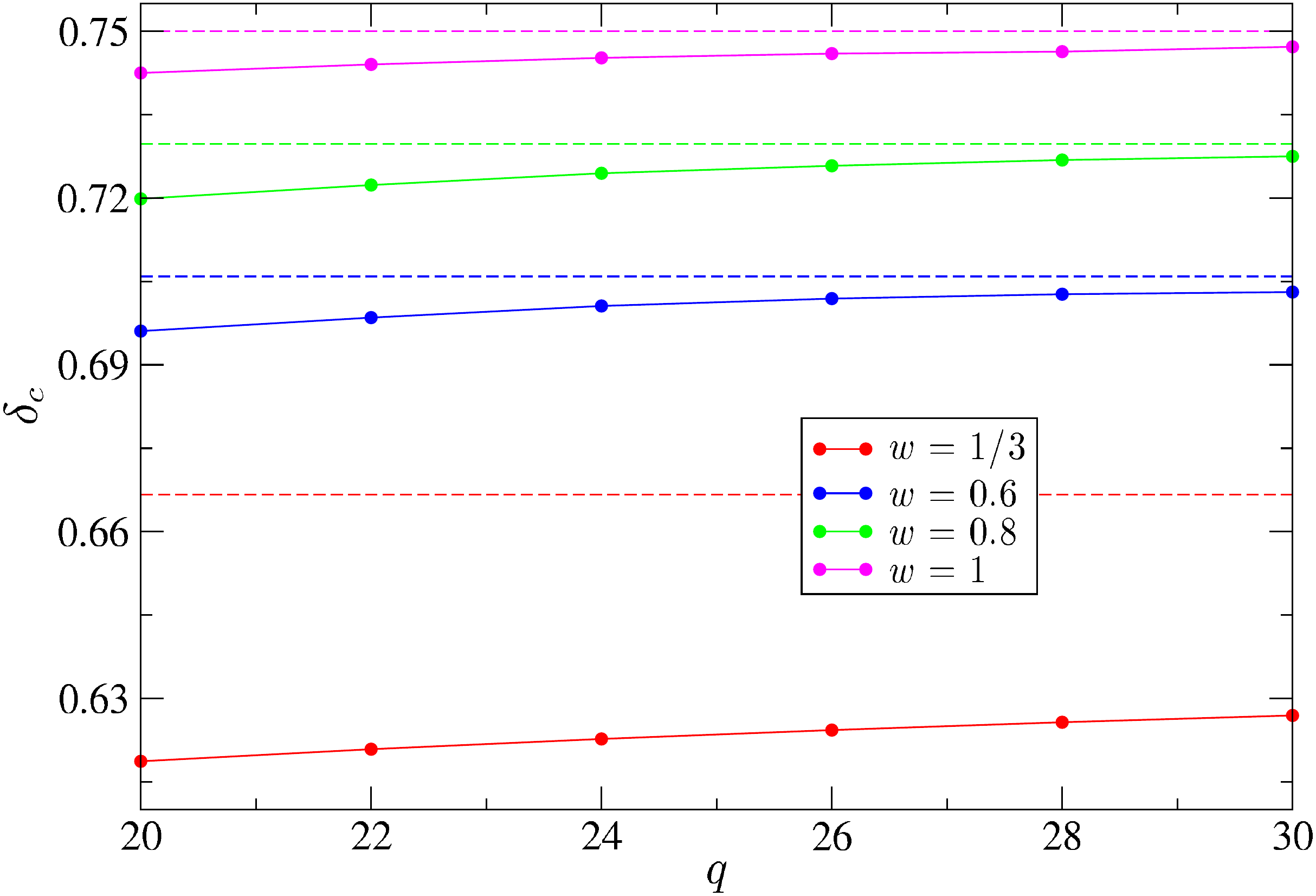} 
  \includegraphics[width=0.45\linewidth]{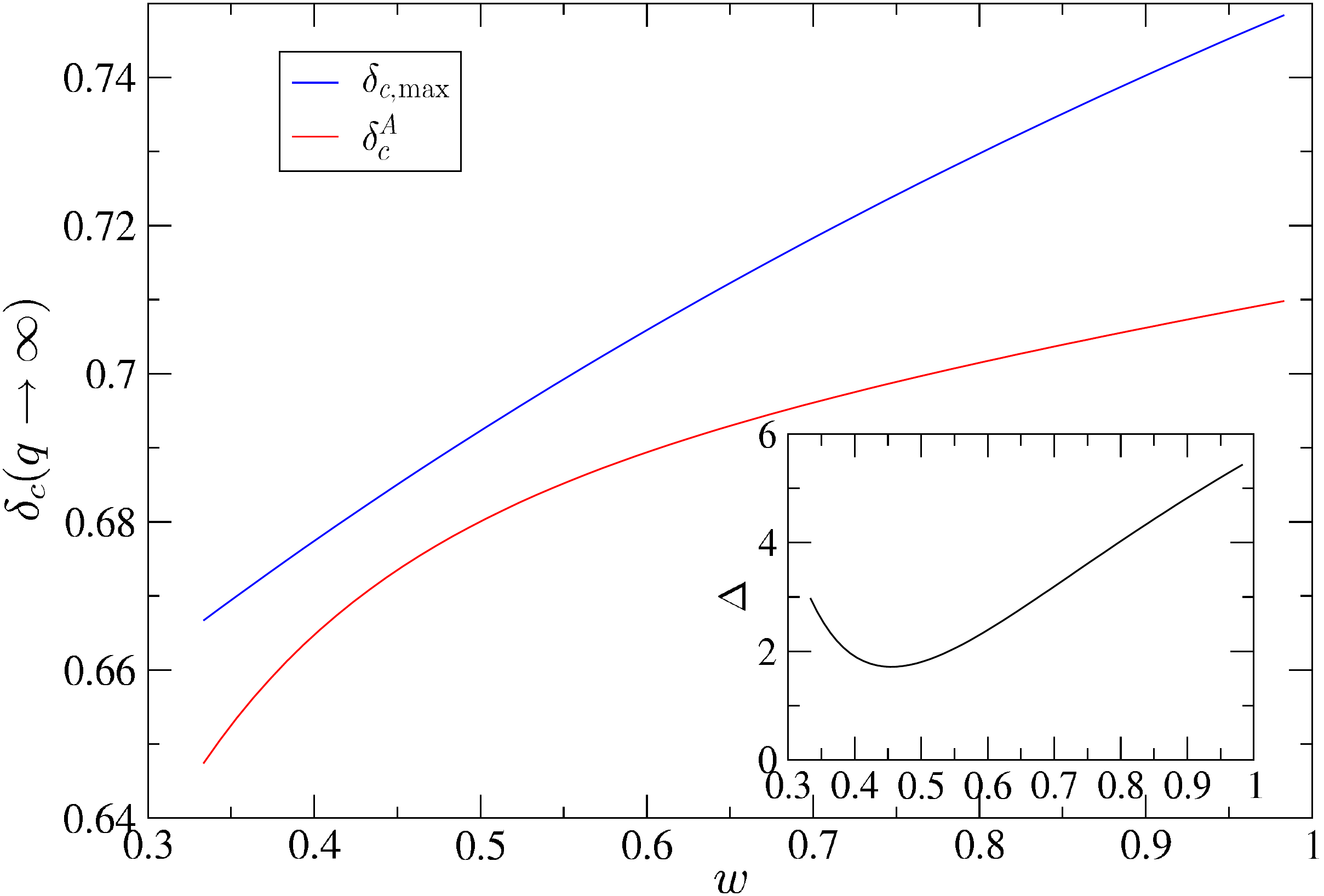} 
  \caption{The $q\gg 1$ limit.  Left: For each $w$ (as labeled), the critical threshold measured in simulations $\delta_c^N$ (symbols connected by solid lines) approaches $\delta_{c,\rm max}\equiv f(w)$ of Eq.\eqref{eq:fw} (dashed) as $q$ increases.  Right: Comparison of the maximum threshold $\delta_{c,\rm max}=f(w)$ and our Eq.\eqref{threshold-anal} when $q \rightarrow \infty$. Inset shows the percent difference between the two.}
  \label{fig:bigq}
\end{figure}

It is easy to show analytically that the compaction function cannot exceed $f(w)$ \cite{ilia}.  Moreover, numerical simulations of $w=1/3$ show that this limit is saturated when the compaction function is sharply peaked \cite{ilia}. Sharply peaked implies $q\to\infty$: for such profiles the pressure gradients fighting the collapse are maximal and thus the compaction function should be too.  This saturation should persist to larger values of $w$ because larger values of $w$ also imply larger pressures which fight the collapse. Therefore, for $w\geq 1/3$ the compaction function of a peaked profile must also saturate the bound.  The left hand panel of Fig. \ref{fig:bigq} shows that, indeed, for $w\geq 1/3$ $\delta_c\rightarrow f(w)$ when $q\rightarrow\infty$ (the case for $\omega=1/3$ was already reported in \cite{ilia}). Therefore, it is interesting to ask how well our Eq.\eqref{threshold-anal} does if we continue to use it even for $q\gg 30$.  The right hand panel of Fig.~\ref{fig:bigq} shows that setting $q\to\infty$ in Eq.\eqref{threshold-anal} returns $\delta_c$ that is within 5\% of $f(w)$ for all $w>1/3$.  This strongly suggests that one can use it for all $q>0.1$.

\subsection{The $q\ll 1$ limit}
\begin{figure}
  \centering
  \includegraphics[width=0.45\linewidth]{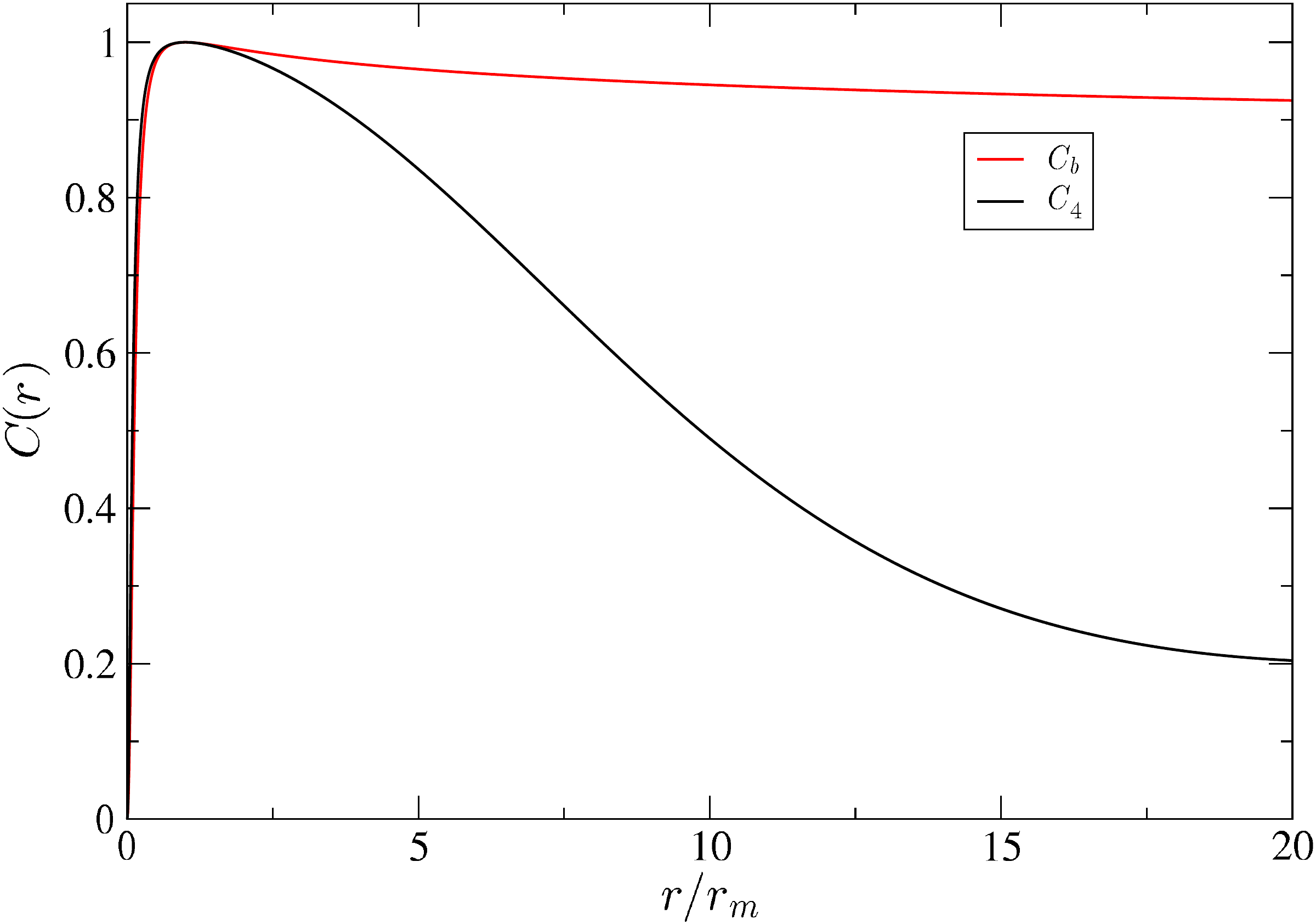} 
  \includegraphics[width=0.45\linewidth]{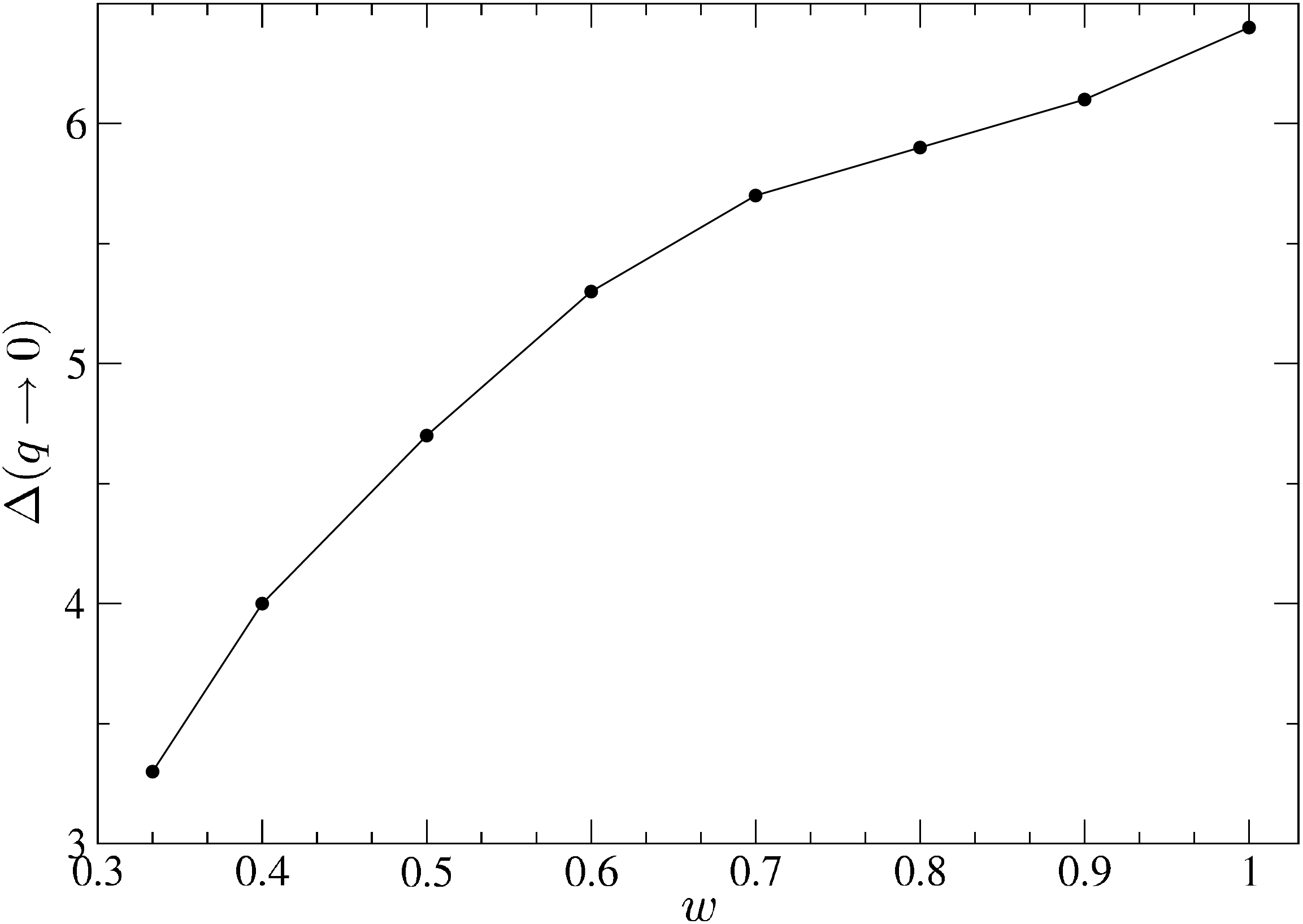} 
  \includegraphics[width=0.45\linewidth]{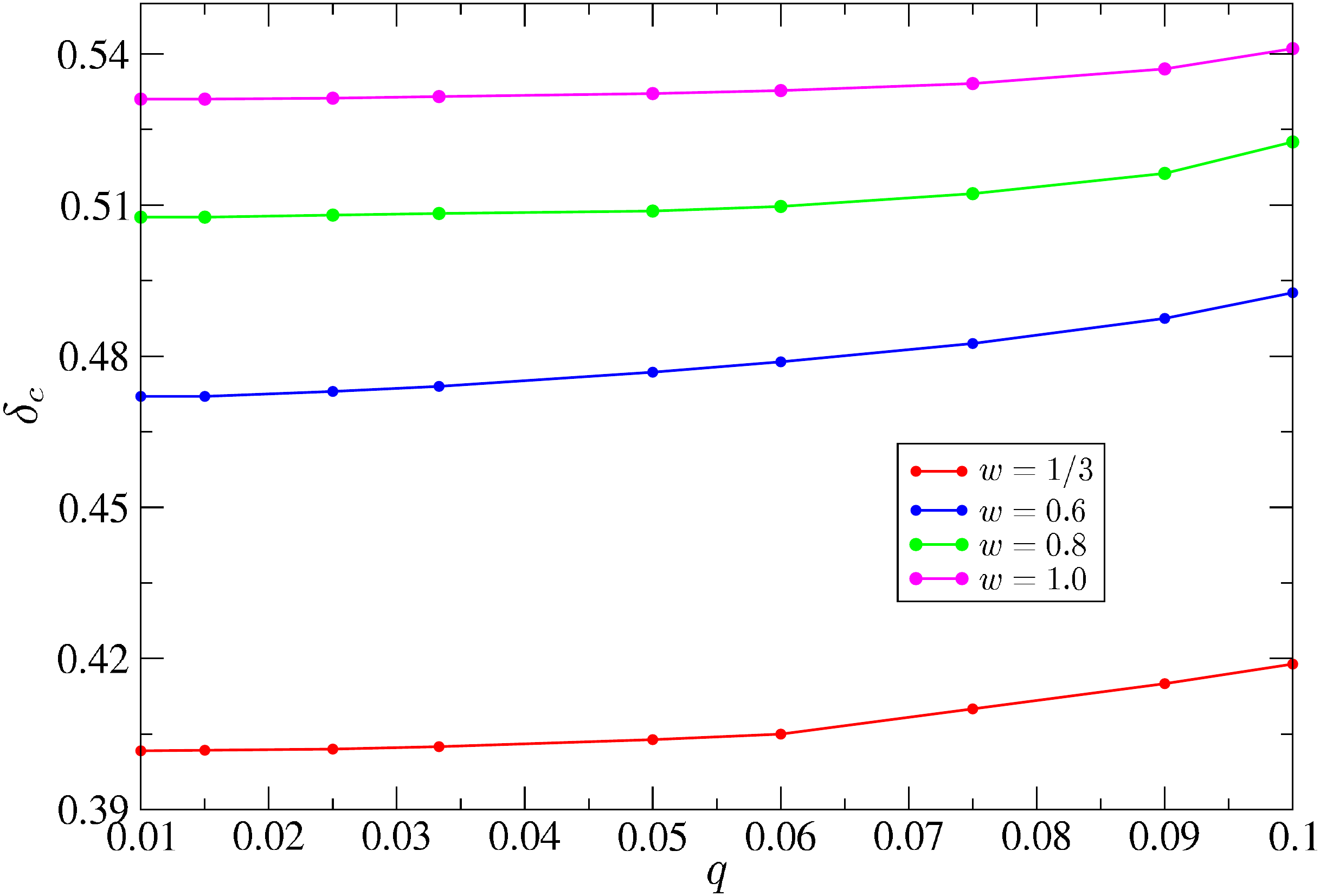} 
  \includegraphics[width=0.45\linewidth]{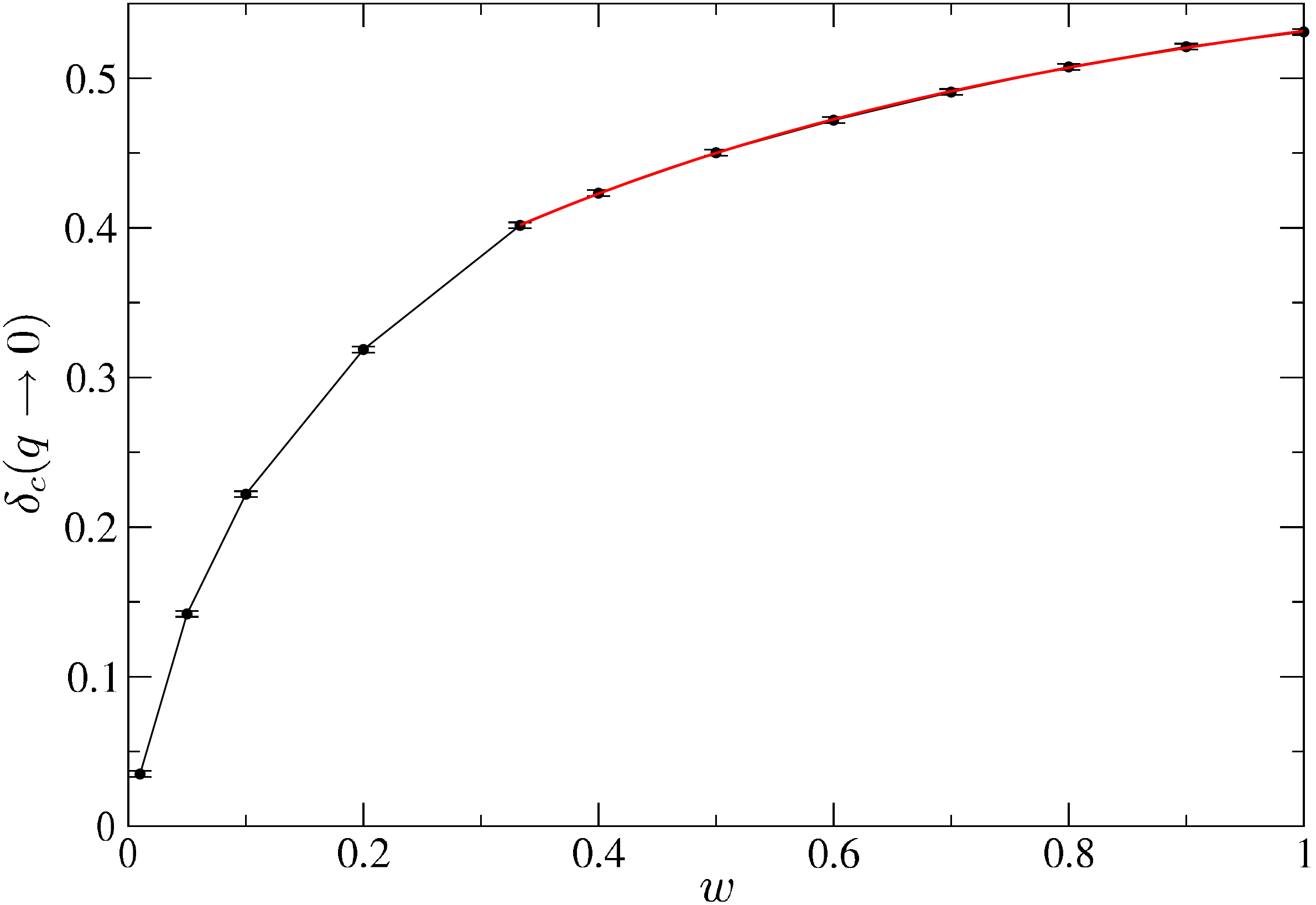} 
  \caption{The $q\to 0$ limit.  Top left: Comparison of compaction functions associated with two profiles having $q=0.015$:  $\com_4$ has $(r_{m,1},r_{m,2},q_{1},q_{2},\gamma) = (150,800,0.0005,0.3,-0.8)$ in Eq.\eqref{eq:LC} and $\com_{\rm b}$ has $(r_m,q)=(37.27,0.015)$ in Eq.\eqref{basis_pol}.  Although they are similar at $r<r_m$, $\com_4$ is much smaller at $r\gg r_m$, so it is easier to simulate accurately.  Bottom left: Convergence of the threshold $\delta_{c}(q,w)$ for profiles of the form Eq.\eqref{eq:LC} to its $q=0$ value, for different $w$ (as labeled).   Bottom right: Numerical threshold for the case $q\to 0$ (symbols and black line) and the result of setting $q\to 0$ in the fitting formula of Eq.\eqref{eq:average_c_new} (red line). Top right: Percent difference between the $q\to 0$ limit of our analytical threshold Eq.\eqref{threshold-anal} and the threshold obtained from simulations of $\com_4$ as $q\to 0$.}
  \label{fig:smallq}
\end{figure}

We now consider $q<0.1$, for which $\com_{\rm b}$ becomes approximately constant over a wide range of scales, making it difficult to simulate the $q\to 0$ limit.  The top left panel of Fig.~\ref{fig:smallq} shows why this limit is better studied by simulating the evolution of profiles given by $K_4$ rather than $K_{\rm b}$.  The two curves show profiles that both have $q=0.015$; however, $\com_4$ is obviously smaller at $r\gg r_m$.  In particular, $\com_4$ satisfies the condition of a flat FRW universe at the boundary much better than does $\com_{\rm b}$.

We have used the $K_4$ profiles to study $\delta_c$ as $q\to 0$.  The bottom left panel of Fig.\ref{fig:smallq} shows that, for all $w>1/3$, $\delta_c$ has approximately converged to its $q\to 0$ value even when $q\sim 0.015$.  The symbols in the right hand panel show that $\delta_c$ in the $q\to 0$ limit is a strong function of $w$. The red curve shows that this dependence is well described by the $q\to 0$ limit of our Eq.\eqref{threshold-anal}, even though Eq.\eqref{threshold-anal} was only calibrated over the range $q\in [0.1,30]$.  Finally, the top right panel shows that the difference between the $q\to 0$ limit of our Eq.\eqref{threshold-anal} and the $q\to 0$ threshold in our simulations of $\com_4$ profiles is typically smaller than about 6 percent.

In summary: We have shown that, as was true for equations of state having $w=1/3$ \cite{RGE}, 
\begin{itemize}	
 \item i) the critical threshold for PBH formation depends mainly on the shape of the compaction function around its peak;
 \item ii) the average of the compaction function over an appropriately chosen volume is a nearly universal quantity which only depends on $w$;
 \item iii) the critical threshold saturates to the maximum of the compaction function in the limit $q\rightarrow \infty$;
 \item iv) for small values of $q$, $\delta_c(q,w)$ rapidly converges to a $q$ independent function,
\end{itemize}   
for all $w \in [1/3,1]$.

\subsection{Combining the two limits to build a fully analytic approach}\label{sec:compare_anal}
One of the steps in our methodology was the assumption that the dependence of the averaged critical compaction function on profile shape is weak enough to be ignored (Eq.\ref{assumption}).  With this in mind, we have explored what happens if, instead of performing a numerical minimization to determine $\alpha(w)$ and $\bar\com_{\rm c}(w)$, we use either the $q\to\infty$ or the $q=0$ limiting values as the basis for our method.  The $q\to 0$ limit has constant $\com$, so $\bar\com_{\rm c}(w,q\to 0) = \delta_c(w,q\to 0)$.  The $q\to\infty$ limit has $K$ (rather than $\com$) $\to$ constant for $r\le r_m$.  Since this limit has $\delta_c\to f(w)$, it has  
\begin{equation}
  \bar\com_{\rm c}(w,q\to\infty)
  = f(w)\frac{3}{5}\frac{1-[1-\alpha(w)]^5}{V[\alpha(w)]}.
 \label{eq:largeq_sol}
\end{equation}
I.e., in these two limits $\bar\com_c$ is not an arbitrary function of $w$.  

\begin{figure}[t]
  \centering
  \includegraphics[width=0.4\linewidth]{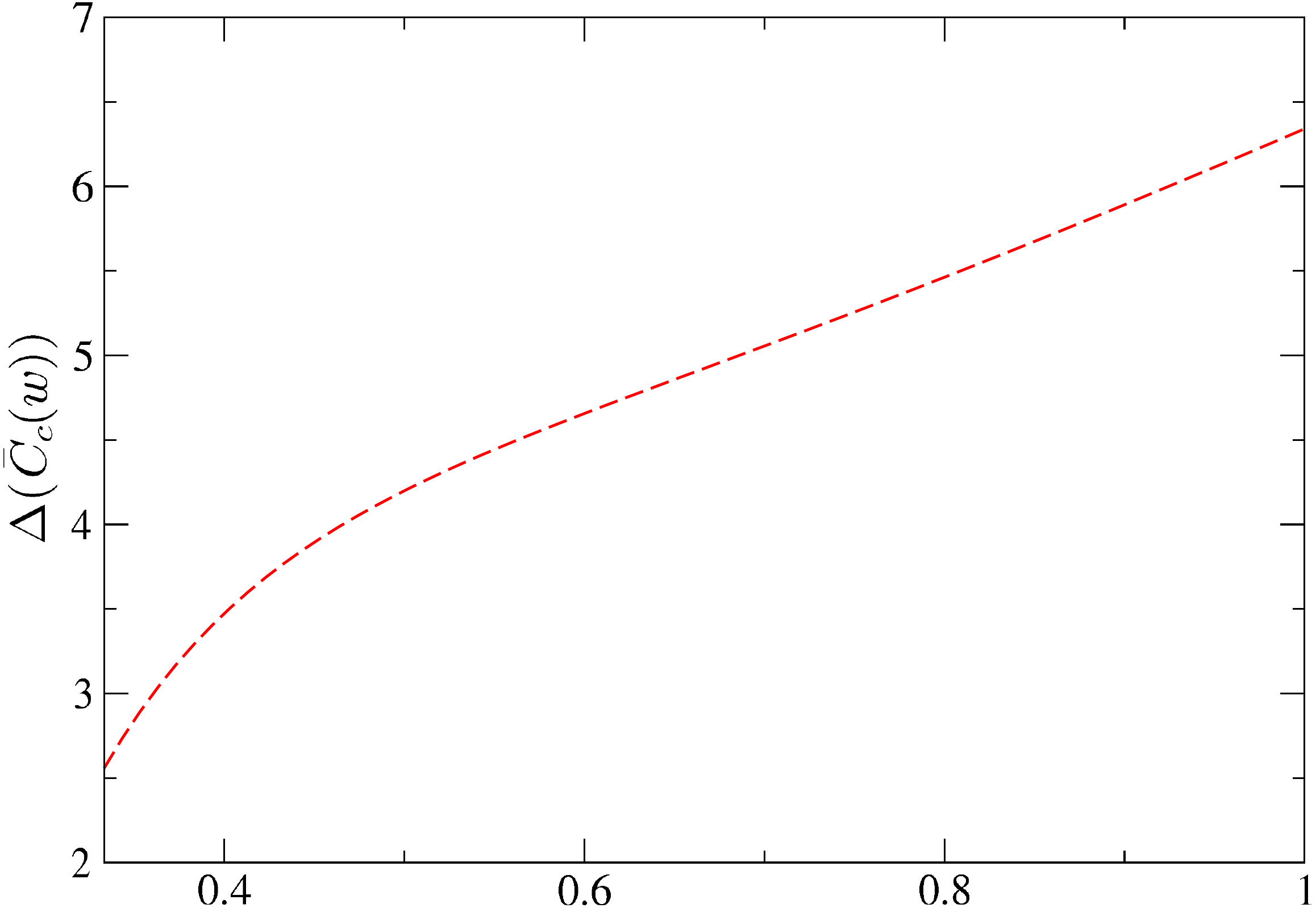} 
  \caption{Percent difference between two estimates of $\bar\com_{\rm c}(w)$:  Eq.\eqref{eq:average_c_new} (which equals $\delta_c(w,q\to 0)$ shown in the bottom right panel of Fig.\ref{fig:smallq}) and Eq.\eqref{eq:average_c}.}
  \label{fig:barCw}
\end{figure}

With this in mind, we start by using the fact that the simulated values of $\delta_{c}(w,q \rightarrow 0)$ directly determine $\bar\com_{\rm c}(w)$.  We have found that the dependence on $w$ (c.f. the bottom right panel of Fig.~\ref{fig:smallq}) is well approximated by
\begin{equation}
 \label{eq:average_c_new}
 \bar\com_{\rm c}(w) = i + j \, {\rm Arctan}(p\, w^l) , \qquad {\rm with}\qquad
 (i,j,p,l) =  (0.262285, 0.251647, 1.82834, 0.984928).
\end{equation}

\begin{figure}[b]
\centering
  \includegraphics[width=0.425\linewidth]{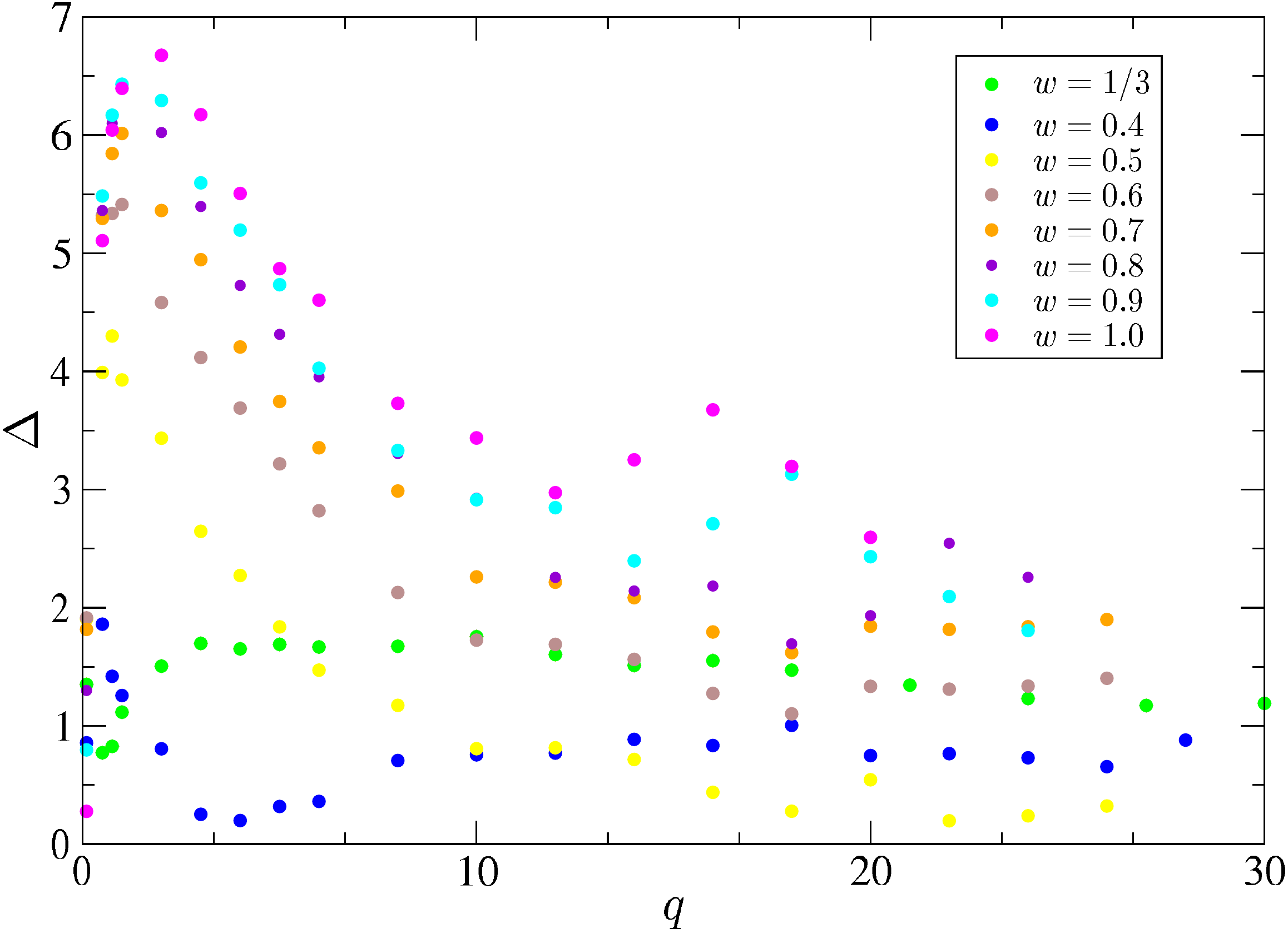} 
  \includegraphics[width=0.425\linewidth]{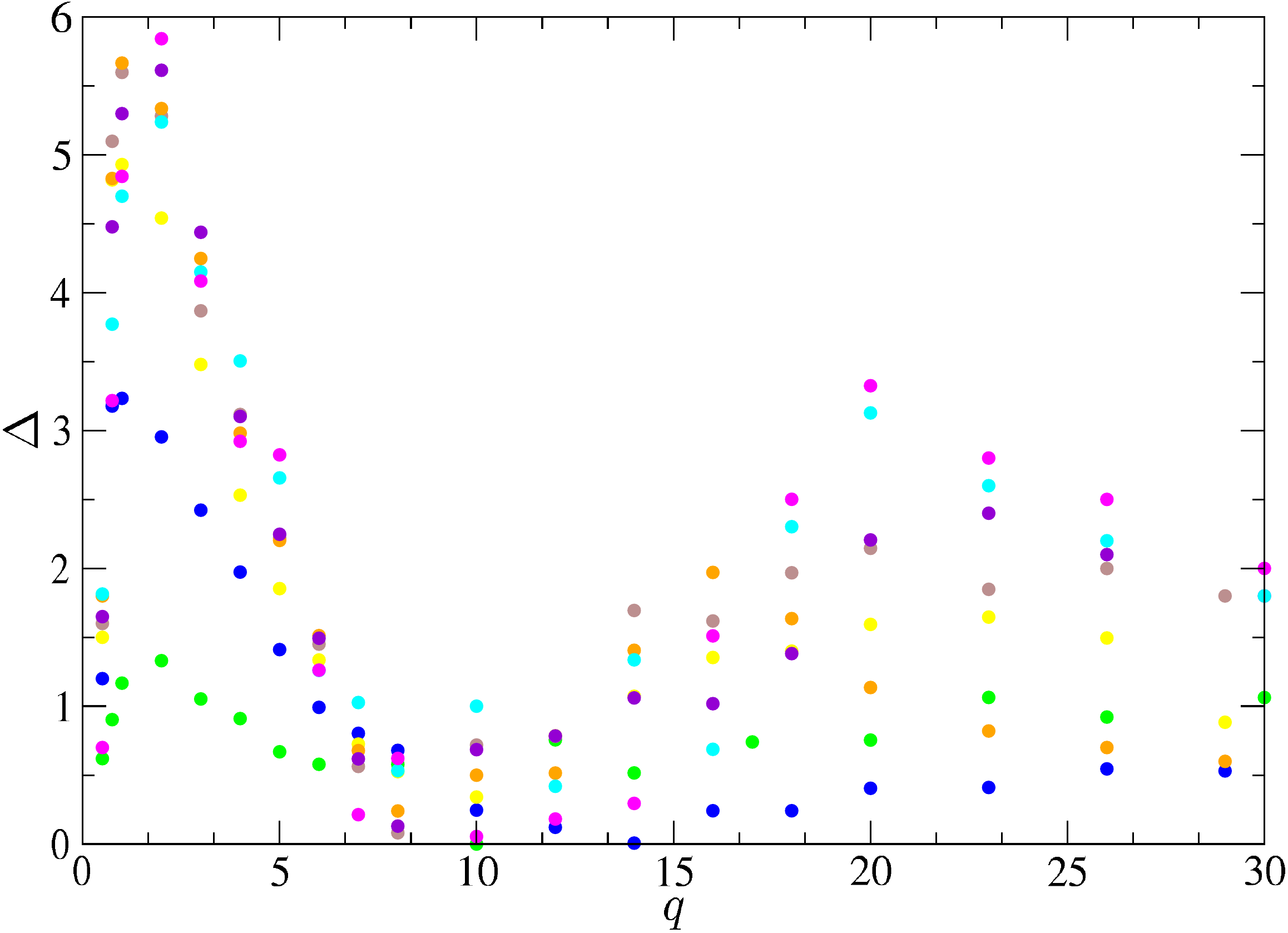} 
  \includegraphics[width=0.425\linewidth]{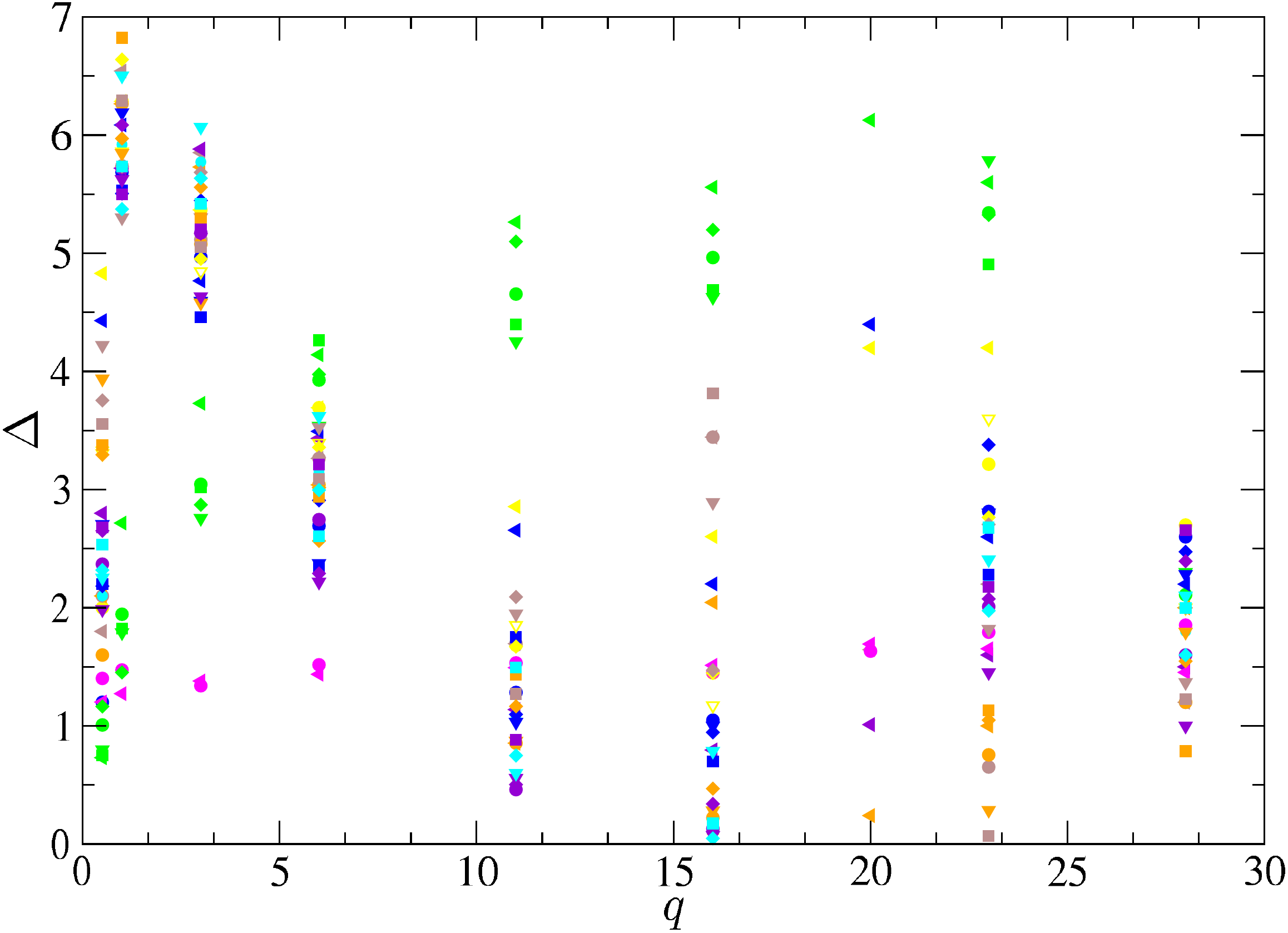} 
\includegraphics[width=0.425\linewidth]{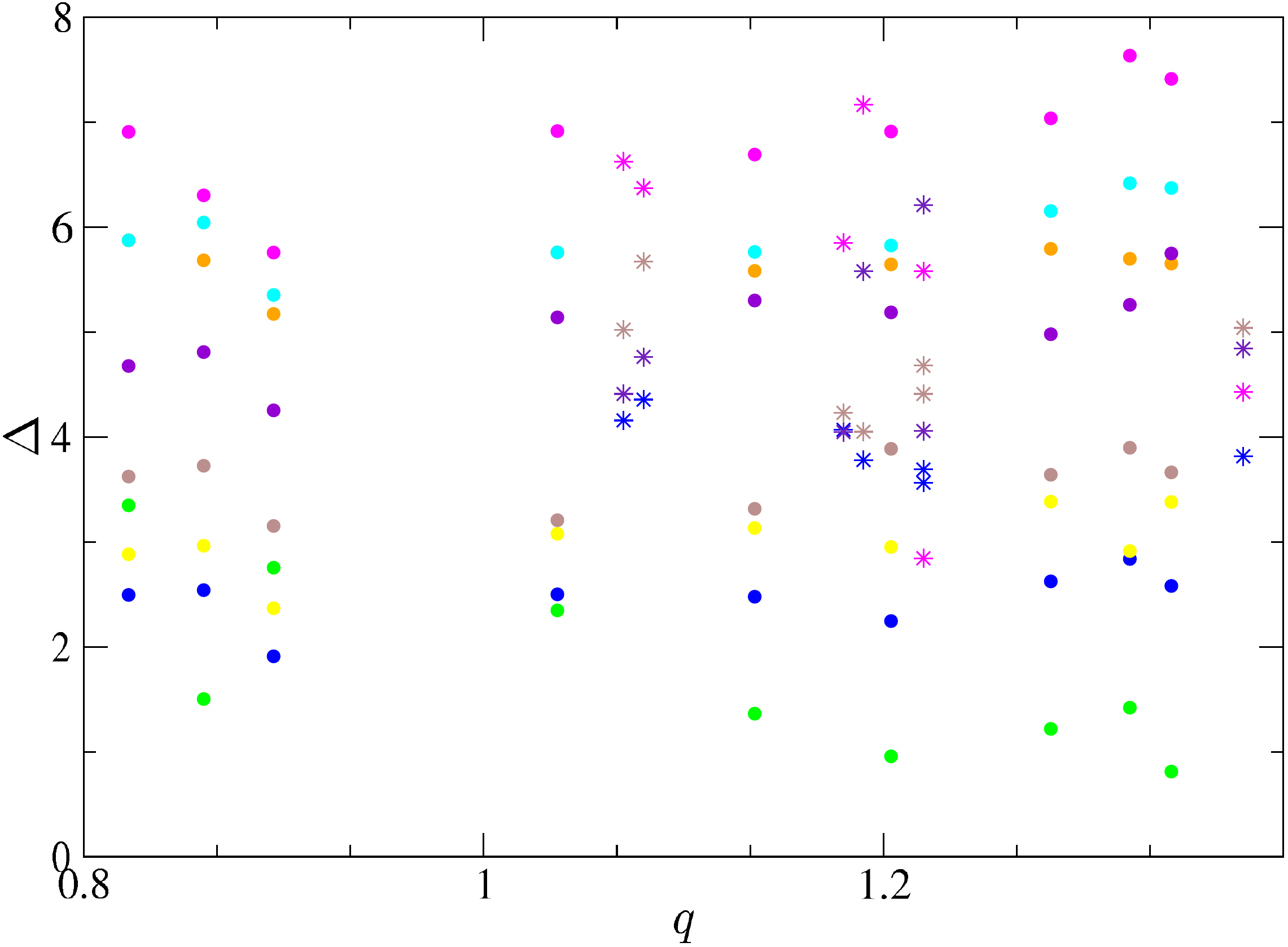} 
\caption{Same as Fig.\ref{fig:dev}, but now the analytic values $\delta^{A}_{c}$ come from using the new fits for $\bar\com_{\rm c}$ and $\alpha(w)$ (Eqs.\ref{eq:average_c_new} and~\ref{eq:alpha_new} instead of Eqs.\ref{eq:average_c} and~\ref{eq:alpha}) in Eq.\eqref{threshold-anal}.
}
\label{fig:dev_new}
\end{figure} 

Fig.~\ref{fig:barCw} shows that this expression for $\bar\com_c$ and that given by Eq.\eqref{eq:average_c} agree to better than 7 percent.  Next, by requiring this $\bar\com_{\rm c}(w)$ to match Eq.\eqref{eq:largeq_sol} we determine $\alpha(w)$, which we have found is well described by 
\begin{equation}
 \label{eq:alpha_new}
 \alpha(w) = m + t \, {\rm Arctan}(r\, w^s) , \qquad {\rm with}\qquad
 (m,t,r,s) = (25261.6, -16081.8, 363647, 2.09818).
\end{equation}
We can now insert Eqs.\eqref{eq:average_c_new} and~\eqref{eq:alpha_new} (instead of Eqs.\ref{eq:average_c} and~\ref{eq:alpha}) in Eq.\eqref{threshold-anal} to produce an analytic estimate of the critical threshold $\delta_c(w,q)$.  Fig.\ref{fig:dev_new} shows the percent difference between these new estimates and the simulated thresholds for a variety of $w$, $q$ and choice of profile family.  Notice that the differences here are not much worse than in Fig.~\ref{fig:dev}, suggesting than if we had an analytic understanding of $\delta_c(w,q\to 0)$ then our methodology for determining $\delta_c(w,q)$ for any $q>0$ would be fully analytic.

\section{Comparison to previous estimates}\label{sec:compare}
In view of the importance of the $q\to 0$ limit, we now compare our results to earlier attempts that were calibrated to small values of $q$.  One is due to \cite{Carr:1975qj}, who used a Jeans length approximation to argue that
\begin{equation}
 \label{eq:CARR}
 \delta_{\rm Carr}= w\ .
\end{equation}
The other is due to \cite{harada}, who improved on \cite{Carr:1975qj} by considering the collapse of a homogeneous overdense sphere surrounded by a thin underdense shell. \cite{harada} argued that, under certain assumptions on the form of the relativistic Jeans instability,  
\begin{equation}
 \label{eq:HYK}
 \delta_{\rm HYK} = f(w)\,\sin^2\Big(\pi v(w)\Big)\ ,
\end{equation}
where $f(w)$ and $v(w)$ are given by Eqs.\eqref{eq:fw} and~\eqref{velo}.  To account for uncertainty in how to formulate the relativistic Jeans criteria, \cite{harada} also provided upper and lower bounds on $\delta_c$ for each $w$.  These are given by their Eqs.(4.36) and~(4.37).  

\begin{figure}[t]
\centering
\includegraphics[width=0.6\linewidth]{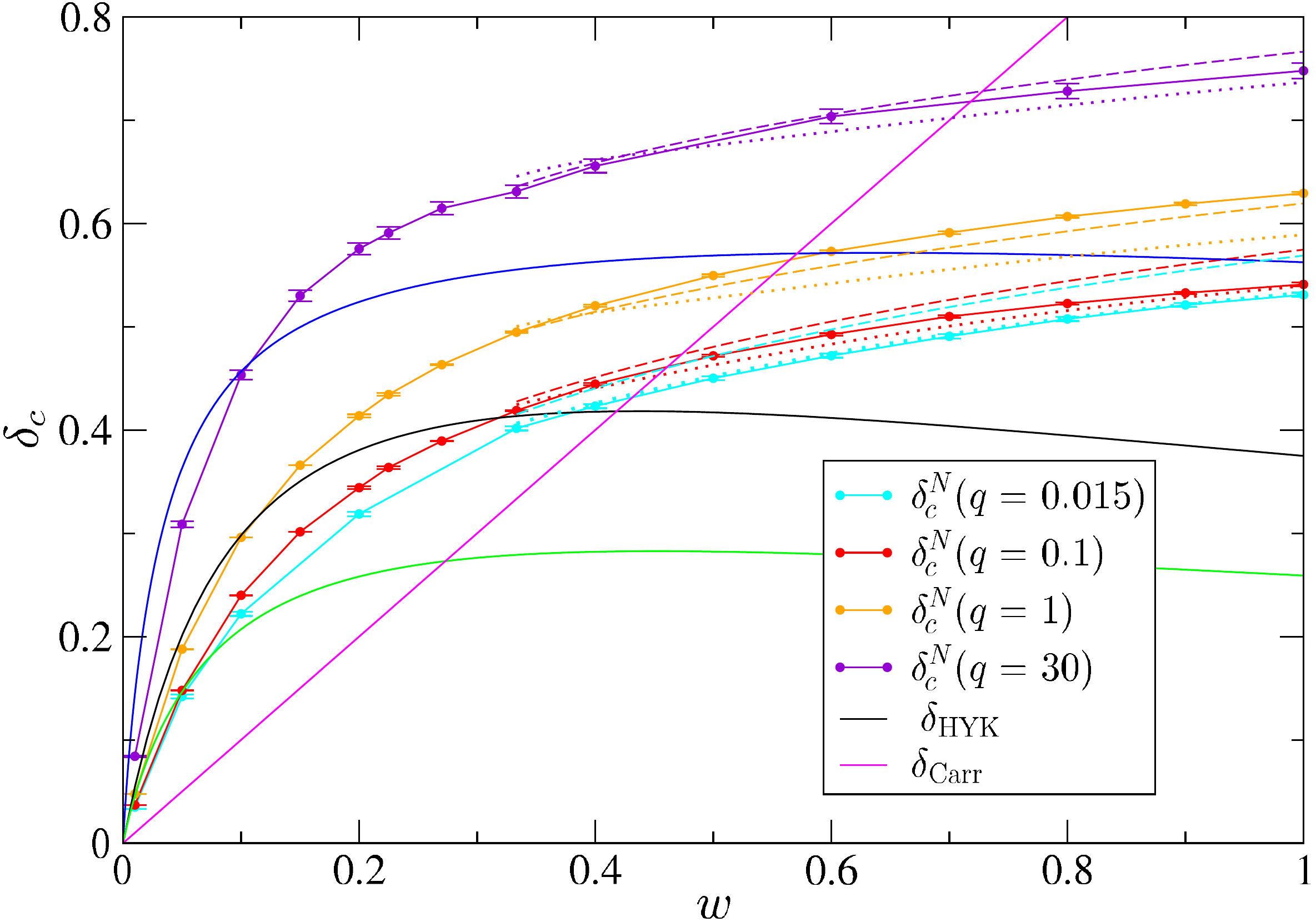} 
\caption{Dependence of threshold $\delta_c$ on $w$ when the initial profile is given by Eq.\eqref{basis_pol} with $q=30$, $q=1$, $q=0.1$ and $q=0.015$ ($q=0$ would be a homogeneous sphere).  Blue and green curves show the maximal and minimal bounds on $\delta_c$ from \cite{harada}.  Solid lines with dots and error bars show the results of our simulations. Magenta line shows the approximation of Carr (our Eq.\ref{eq:CARR}); black curve labeled HYK is from \cite{harada} (our Eq.\ref{eq:HYK}).  Neither predicts $q$ dependence of $\delta_c$, but $\delta_{\rm HYK}$ explicitly aims to describe the $q\ll 1$ limit. The other curves show our approximation (Eq.\ref{threshold-anal}) in which $\delta_c$ depends both on $w$ and $q$. The dotted curves use Eqs.\eqref{eq:average_c_new} and~\eqref{eq:alpha_new} in Eq.\eqref{threshold-anal} whereas the dashed curves use Eqs.\eqref{eq:average_c} and~\eqref{eq:alpha} in Eq.\eqref{threshold-anal}.}
\label{fig:harada}
\end{figure}

Neither Eq.\eqref{eq:CARR} nor~\eqref{eq:HYK} admit dependence on the profile shape, which we showed are present.  Nevertheless, it is interesting to see how well they perform.  The solid lines in Fig.\ref{fig:harada} show these approximations; symbols with error bars show $\delta_c(w)$ from numerical simulations of profiles having $q=0.015$, $q=0.1$, $q=1$ and $q=30$. The dashed and dotted curves, which provide a significantly better description of the simulations, show the result of inserting Eqs.\eqref{eq:average_c} and~\eqref{eq:alpha}, or Eqs.\eqref{eq:average_c_new} and~\eqref{eq:alpha_new}, in our Eq.\eqref{threshold-anal}. In both cases, our Eq.\eqref{threshold-anal}, like the simulations, exceeds even the upper bound claimed by \cite{harada} at ever lower $w$ as $q$ increases.

The discrepancy between our simulations and Eq.\eqref{eq:HYK} at small $w$ -- which is as large as 50\% for $q=0.1$ -- deserves further comment, as this is the limit that was believed to be optimal for the approximations on which Eq.\eqref{eq:HYK} is based.  This discrepancy is even larger than the one noticed earlier because \cite{harada} only compared their formula with simulations of a Gaussian curvature profile (i.e., Eq.\ref{eq:exp} with $q=1$).  Indeed, for $w<0.15$ the solid black curve {\em does} provide a reasonable description of our $q=1$ simulations (even though the profile is given by Eq.\ref{basis_pol} rather than Eq.\ref{eq:exp}, so it is not exactly Gaussian in shape).  However, the top-hat profile, which is the one used in the  analytic calculations of \cite{harada}, is much better approximated by $q\ll 1$.  For $q=0.1$, their formula does not describe the simulations particularly well, and the discrepancy at $w<1/3$ is even worse when $q=0.015$.  This disagreement suggests that the apparent agreement shown in Fig.3 of \cite{harada} is just a result of numerical coincidences:  it is not physical.  Therefore, analytic understanding of $\delta_c$ in the $q\to 0$ limit remains an open and -- our analysis suggests -- extremely interesting and impactful problem.

\section{Conclusions}

We performed numerical simulations of black hole formation from spherically symmetric super-Hubble perturbations in a cosmological background that is a perfect fluid having equation of state $p=w\rho$ with $w\in (0,1]$.  The simulations use pseudospectral methods \cite{albert_paper}, and generalize our previous study of the case $w=1/3$ \cite{RGE} to other $w$ with similar reliability and accuracy (Figs.~\ref{fig:constraint1}--\ref{fig:constraint4} and Appendix~\ref{sec:testsims}).  The simulations show that, for a black hole to form, the compaction function $\com$ (Eq.\ref{compactionfunction}) must exceed a critical threshold $\delta_c$.  This $\delta_c$ depends on $w$ and on the `shape' -- the radial profile -- of the perturbation (Figs.~\ref{fig:fitting_basis} and~\ref{fig:harada}).

  We argued that, for $w>1/3$, pressure gradients are strong and erase small scale details of perturbations, so a simple parametrization in terms of suitably chosen averaged quantities should be sufficient to predict $\delta_c$ quite accurately (Sec.~\ref{sec:heuristics}).  We then argued that $w$ determines the scale over which one should average $\com$ and that, given $w$, the shape-dependence of $\delta_c$ can be parametrized using only a single additional parameter, $q$ of Eq.\eqref{q}, which is a dimensionless measure of the curvature of $\com$ on the scale where $d\com/dr=0$.  We demonstrated the accuracy of this proposal using a wide variety of parametrizations of possible profile shapes (Eq.\ref{basis_pol}, Eqs.\ref{eq:exp}--\ref{eq:spectrum} and Fig.~\ref{fig:C_tested}).  Our `universal' formula, $\delta_c(q,w)$ of Eq.\eqref{threshold-anal}, is always within $\sim 6\%$ of the simulated values to which is has been calibrated:  $w>1/3$ (Figs.~\ref{fig:dev_numerical_2}--\ref{fig:smallq} and~\ref{fig:dev_new}).

We also showed that the expressions for $\delta_c$ provided by \cite{harada}, which are supposed to apply in the limits of small $q$ and $w$, are not as accurate as our Eq.\eqref{threshold-anal} (Fig.~\ref{fig:harada}). A full analytic understanding of the $q\to 0$ limit, if it exists, remains a very interesting problem that would make our semi-analytic work fully analytic (c.f. Sec.\ref{sec:compare_anal}). Nevertheless, even without this understanding, our $\delta_c(q,w)$ of Eq.\eqref{threshold-anal} is sufficiently accurate that it vastly simplifies estimates of PBH abundances when the equation of state has $w\geq 1/3$. Indeed, our semi-analytical formula for the PBH threshold formation, removes the need to numerically simulate the evolution of every single profile shape that is statistically likely. 

The fact that this spherically symmetric case has worked out so easily, at least for $w>1/3$, suggests that a number of other problems may also be tractable.  For example, PBH formation from non-spherical perturbations \cite{elipsoidal,yoolate}, the effects of rotation on the gravitational collapse \cite{rotating_PBH,carsten,riotto} or even PBH formation in modified gravity models \cite{chen2019threshold} are all interesting directions for future work which our analysis enables.  Finally, how to make progress when $w<1/3$ is another open question.  
  
\appendix

\section{Convergence tests of the numerical simulations}\label{sec:testsims}

To check the reliability of our simulations, we have performed a similar test to the one described in Ref.\cite{albert_paper}.  The initial conditions of our simulations rely on the gradient expansion approximation. While the evolution equations are kept un-altered, this approximation slightly violates the Hamiltonian constraint which is the derivative of the mass definition eq. \eqref{massdef} i.e. $M'\equiv 4\pi R' R^2 \rho$. 

Defining then
\be
\Psi=\frac{M'_{\rm num}-M'_{\rm def}}{M'_{\rm def}}=\frac{M'_{\rm num}/R'_{\rm num}}{4\pi\rho_{\rm num}R^2_{\rm num}}-1\ ,
\ee 
the numerical square norm 
\begin{equation} 
 \mid\mid \Psi \mid\mid_{2} \equiv \frac{1}{N_{\rm cheb}}\sqrt{\sum_{k} \Big|  \frac{M_{k}'/R_k'}{4 \pi \rho_{k} R_{k}^{2}}-1 \Big|^2},
\label{eq:constraint}
\end{equation}
where $k$ labels each point of the grid $x_{k}$, should be much smaller than $1$ for a self-consistent evolution.

\begin{figure}
\centering
  \includegraphics[width=0.75\linewidth]{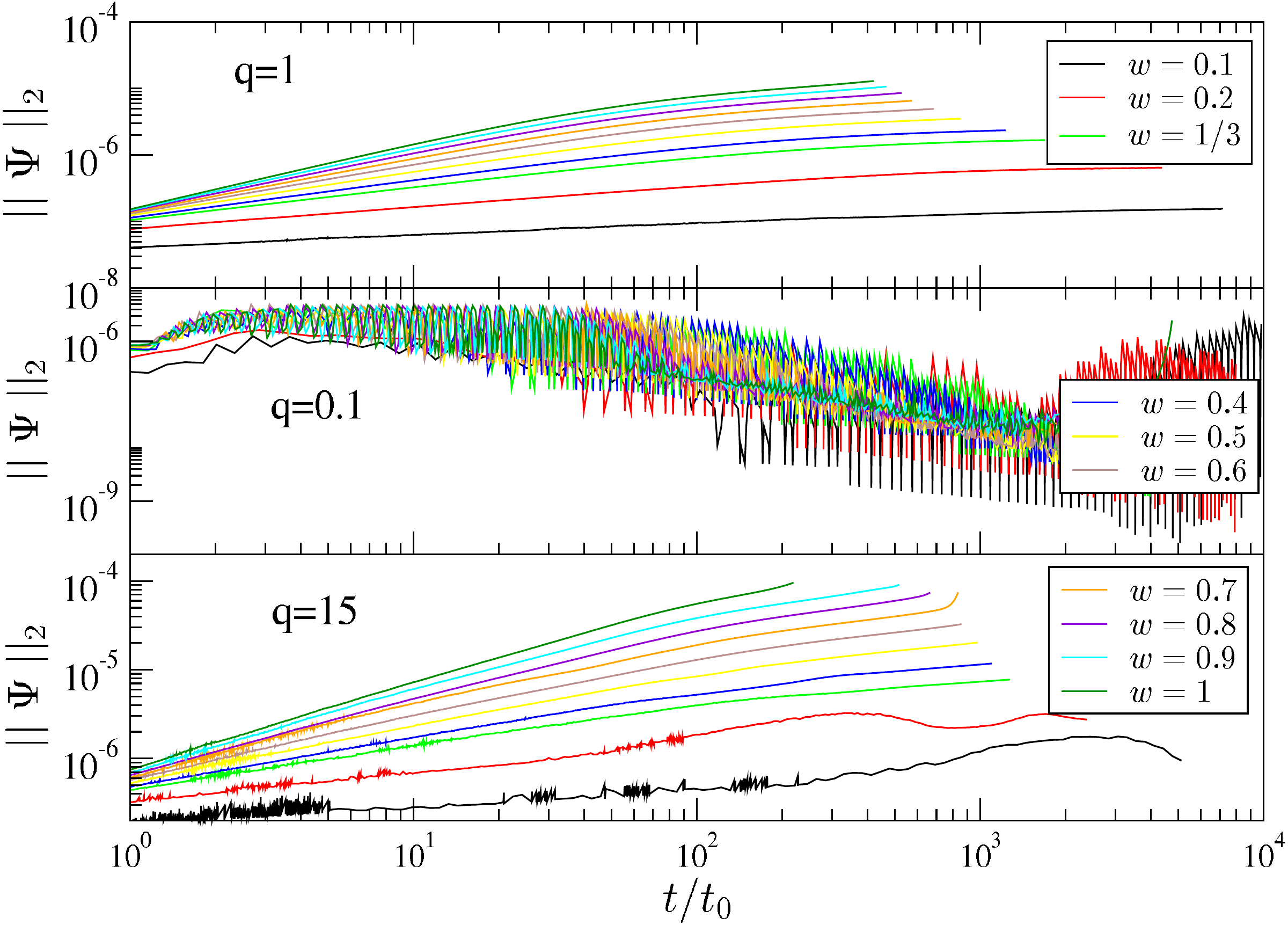} 
  \caption{Evolution of the Hamiltonian constraint (Eq.\ref{eq:constraint}) using our fiducial profile choice, Eq.~\eqref{basis_pol}, for different values of $q$ (top to bottom) and $w$ (as labeled) when $\delta = \delta_c(w,q) + 10^{-2}$ is supercritical. The same qualitative behavior is obtained for subcritical $\delta$ (i.e. $\delta<\delta_{c}(w,q)$).}
\label{fig:constraint1}
\end{figure}

\begin{figure}
\centering
  \includegraphics[width=0.75\linewidth]{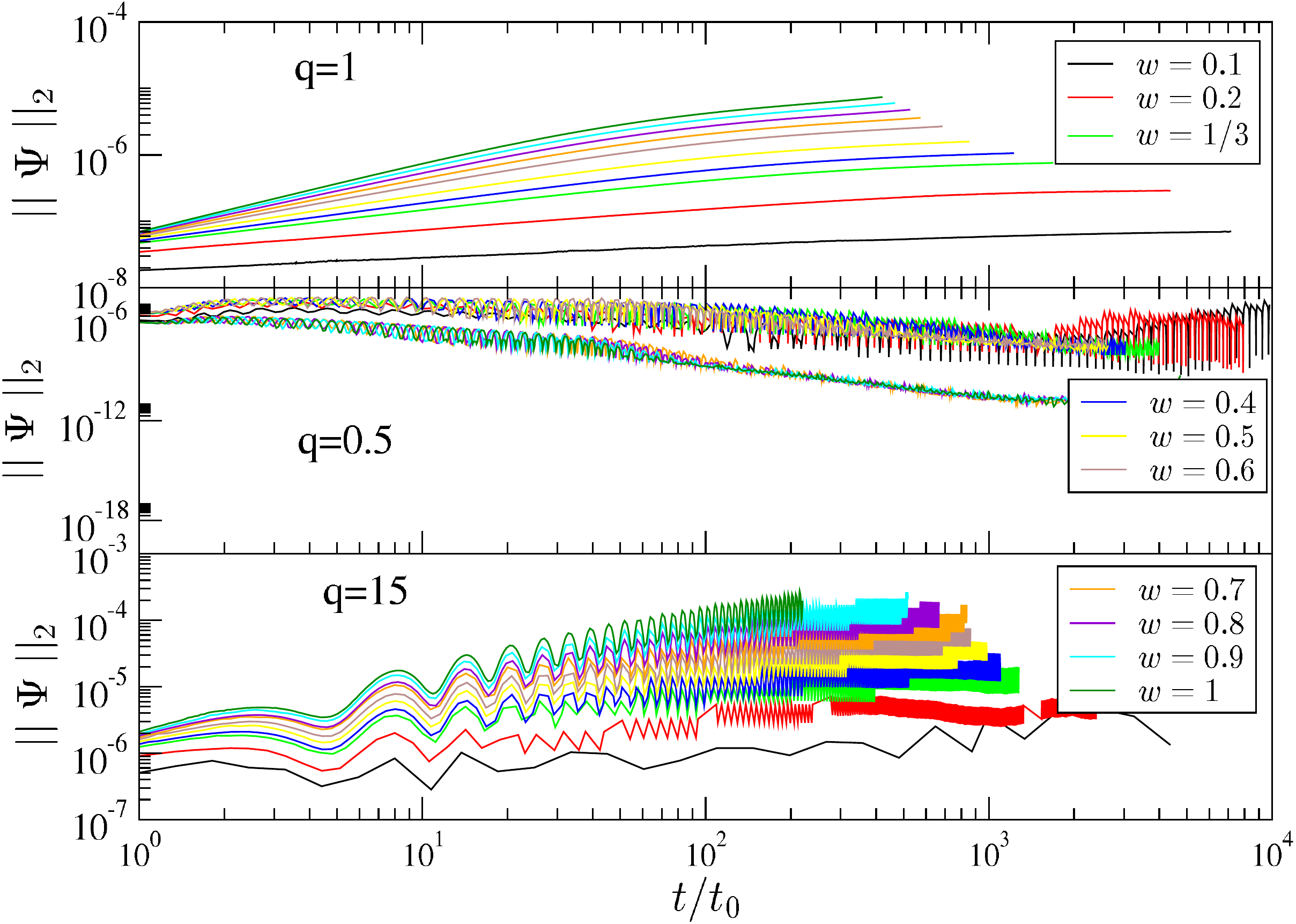} 
  \caption{Same as Fig.~\ref{fig:constraint1} but using Eq.~\eqref{eq:exp} for the profile shape.}
\label{fig:constraint2}
\end{figure}

\begin{figure}
\centering
  \includegraphics[width=0.75\linewidth]{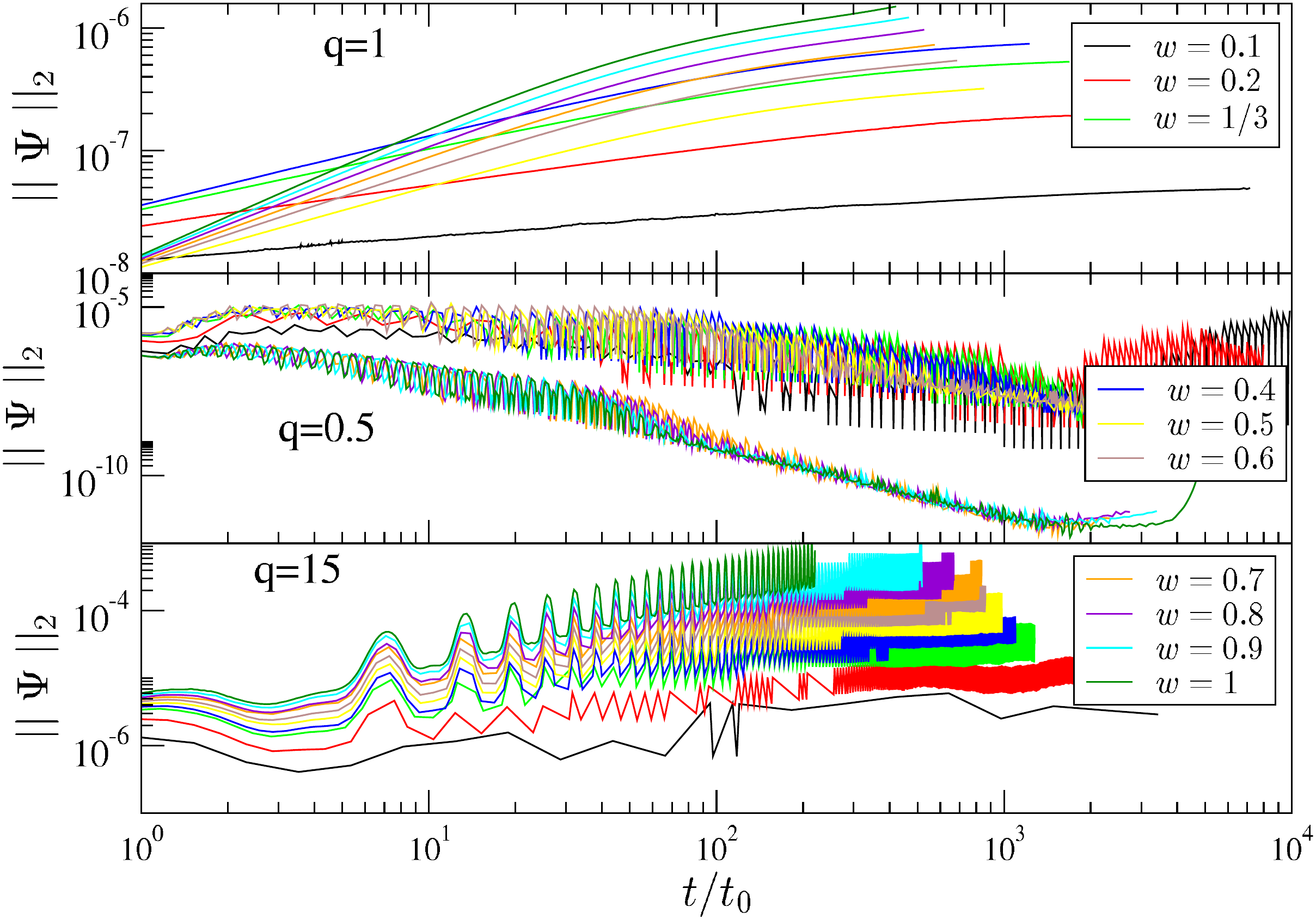} 
  \caption{Same as Fig.~\ref{fig:constraint1} but using Eq.~\eqref{eq:lamda} with $\lambda=1$ for the profile shape.}
\label{fig:constraint3}
\end{figure}

\begin{figure}
\centering
  \includegraphics[width=0.75\linewidth]{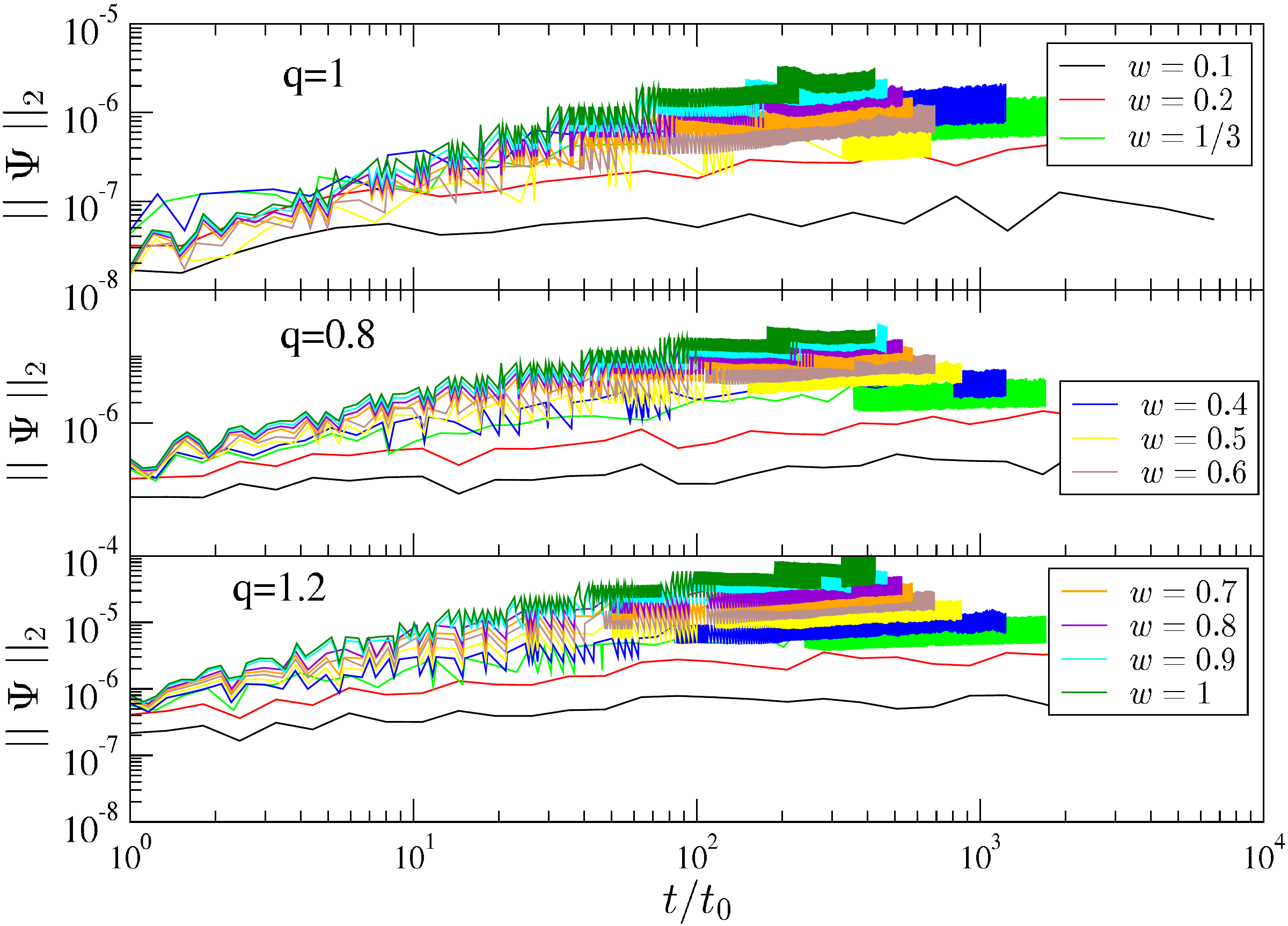} 
  \caption{Same as Fig.~\ref{fig:constraint1} but using Eq.~\eqref{eq:spectrum}  for the profile shape. Note that the range of $q$ here is smaller than for previous figures of this appendix.}
    
\label{fig:constraint4}
\end{figure}

Fig.2 of Ref.\cite{albert_paper} shows that this is indeed the case for a number of profile choices when $w=1/3$.  Figs.\ref{fig:constraint1}--\ref{fig:constraint4} of this Appendix show that this remains true for all $w$ of interest in this paper, and for all the profile shapes and families we have tested.  Roughly speaking, for $q\ge 1$ convergence is more difficult as $w$ increases, but our simulations always have $\mid\mid \Psi \mid\mid_{2}< 10^{-4}$.  

In passing, we also note that the fiducial profiles Eq.\ref{basis_pol}, give more stable numerical evolutions than the basis used in our previous work \cite{RGE}. This is an extra justification that the choice of basis used in this paper is optimal for the reliability of our semi-analytical formulae for the threshold $\delta_c(w,q)$.

\begin{acknowledgments}
CG was supported by the Ramon y Cajal program and by the Unidad de Excelencia Maria de Maeztu Grant No. MDM-2014-0369. AE and CG are supported by the national FPA2016-76005-C2-2-P grants of the Ministerio de Ciencia y Educacion. AE is supported by the Spanish MECD fellowship FPU15/03583.
\end{acknowledgments}

\end{document}